\documentclass[hidelinks,11pt]{article}
\usepackage{arxiv}
\usepackage{cite} 

\usepackage{times}
\usepackage{graphicx}        
\usepackage{multicol}        
\usepackage[bottom]{footmisc}

\usepackage{amssymb,amsmath,mathrsfs,amsthm}
\usepackage{mathtools}

\usepackage{enumerate}
\usepackage[inline,shortlabels]{enumitem} 
\usepackage{framed}
\usepackage{algorithm,algorithmic,psfrag}
\usepackage{soul}
\usepackage{color}
\usepackage{kbordermatrix}
\usepackage[utf8]{inputenc} 
\usepackage[T1]{fontenc}    
\usepackage{hyperref}       
\usepackage{url}            
\usepackage{booktabs}       
\usepackage{amsfonts}       
\usepackage{nicefrac}       
\usepackage{microtype}      
\usepackage{doi}
\usepackage{xifthen}
\usepackage{caption}  
\usepackage[numbered, framed]{matlab-prettifier} 
\usepackage{tcolorbox}  

\newtheorem{ass}{Assumption}

\newtheorem{corollary}{Corollary}
\newtheorem{prop}{Proposition}

\definecolor{paleGreen}{rgb}{.3, .7, .3}
\definecolor{coolBlue}{rgb}{.3, .5, 1}
\definecolor{rosePink}{rgb}{.9, .5, .4}

\newcommand{\matlab}{\mbox{\textsc{Matlab}}}

\newcommand{\DpCoeff}[1]{\alpha_{#1}}
\newcommand{\DqCoeffpr}[1]{\beta_{#1}}
\newcommand{\DqCoeffpl}[1]{\kappa_{#1}}
\newcommand{\DqCoeffql}[1]{\gamma_{#1}}

\newcommand{\R}{\mathbb{R}}

\newcommand{\zero}{\mathbf{0}}
\newcommand{\one}{\mathbf{1}}
\newcommand{\eye}{I}
\DeclareMathOperator*{\diag}{diag}
\newtheorem{defin}{Definition}

\DeclareMathOperator*{\blkdiag}{blkdiag}

\title{A Compendium of Control-Oriented Models of Gas Processing Equipment Components}

\author{Sven Br\"uggemann
\thanks{S. Br\"uggemann and R.R. Bitmead are with the Department of Mechanical \&\ Aerospace Engineering, University of California, San Diego, La Jolla CA 92093-0411, USA (e-mail: \{sbruegge, rbitmead\}@eng.ucsd.edu).} ,
Robert H. Moroto
\thanks{R. H. Moroto was formerly with Solar Turbines {Incorporated}, San Diego CA 92123, USA (e-mail: rhmoroto@gmail.com).} , and 
Robert R. Bitmead,$^*$
\thanks{This work was supported by Solar Turbines {Incorporated}.}
}

\begin{document}
\maketitle

\section*{Introduction}
For transient modeling of gas flow through pipe networks, fluid dynamics and computational fluid dynamics, are well-established subjects focused on high-fidelity modeling given design and boundary conditions. They involve nonlinear partial differential equations (PDEs) and transport phenomena, which are not amenable to finite-dimensional control design but instead are targeted and tested for simulation.

Other more pragmatic modeling for gas pipeline distribution systems \cite{benner2019,kralik1988,Behbahaninejad2008AMS} is usually based on discretization and  yields a system of ordinary differential algebraic equations (DAEs), which again is not well-suited to control design. Although, it can be used directly for controller synthesis in some circumstances \cite{mpdDae} and, as noted in \cite{benner2019}, if the DAE is of index~1. Theorem~4.1 in Benner \text{et al.} \cite{benner2019} establishes that the DAEs describing the gas flow through interconnected pipes are indeed of index~1 and so it is possible to rewrite the DAE as an ordinary differential equation (ODE) without the algebraic constraints. 

This fact is used in \cite{sven_bob_rob_gas} to rewrite the system of DAEs as a (state-space) system of linear ordinary differential equations (ODEs) \textit{subsuming} the algebraic constraints. Thus, to synthesize model-based controllers the rich literature on Linear Systems Theory can be exploited, such as Mason's Gain Formula for modeling interconnections. An equivalent approach for interconnections adapted to state space realizations is presented in \cite{sven_bob_rob_gas}.

This work builds on \cite{sven_bob_rob_gas} and provides linear control-oriented state space models of gas flow through standard equipment, such as valves, compressors, manifolds and non-trivial pipe geometries. After presenting the catalog of components, we show how to interconnect them: we briefly recall the matrix methodology derived in \cite{sven_bob_rob_gas}, accompanied by a Matlab example for a gas loop; secondly for the same example we provide an alternative approach using Matlab's \verb|connect| function.

All the aforementioned models are suited to model-based MIMO control design tools as they are in state-space form and they can be physically parametrized. {Conservation of mass is the source of the algebraic constraints in \cite{benner2019,kralik1988,Behbahaninejad2008AMS} and is a property subsumed in these new models. This is characterized in by model transfer function properties at frequency zero \cite{sven_bob_control}. Each of the component models is shown to posses this property}

{This document is meant to serve as an accompaniment to the material in \cite{sven_bob_rob_gas,sven_bob_control}. Details of assumptions are provided there, along with motivations and derivations from PDEs. Accordingly, it is organized by component type for the first eight sections with Section~9 devoted to the presentation of the approach to interconnection of the components into systems amenable to use with \matlab's control system design tools.}
 
\newpage
\tableofcontents
\newpage
\section*{Notation}
We denote pressures by $\check p(x,t)$ and mass flows by $\check q(x,t)$, where $x$ is the spatial dimension and $t$ the time. For the nominal values we write $\bar p(x)$ and $\bar q(x)$, whereas deviations are described by $p(x,t)=\check p(x,t)-\bar p(x)$ and $q(x,t)=\check q(x,t)-\bar q(x)$. Assume $x\in[0,X]$ for some $X>0$. We define $p_\ell\doteq p(0,t)$ and $p_r \doteq p(X,t)$, and accordingly for the mass flow. {The same convention holds for the temperature $T(x,t)$ where applicable.} Signals related to component $i$ for some $i\in\mathbb{N}$ are denoted by the additional subscript $i$. By~$\zero$ we denote the zero matrix of appropriate size; $\zero_n$ is the related \emph{row vector}, and $\zero_{n,m}$ is the related matrix of dimension $n\times m$. The same notation is used for the matrix of ones, $\one$. For matrix $A$, we write $[A]_{i,j}$ for the matrix element in row $i$ and column $j$, and similarly for vectors. The following table lists the meaning of common parameters.
\begin{table}[ht]
\begin{center}
\begin{tabular}{|c|l|l|}
\hline
Symbol&Meaning&SI-unit\\
\hline\hline
$A$&Cross-sectional area&$\scriptstyle[\text{m}^2]$\\
\hline
$c_v,c_p$&Specific heat capacities&$\scriptstyle[\text{J/(kg K)}]$\\
\hline
$D$&Pipe inside diameter&$\scriptstyle[\text{m}]$\\
\hline
$D_o$&Pipe outside diameter&$\scriptstyle[\text{m}]$\\
\hline
$E$&Total energy&$\scriptstyle[\text{J}]$\\
\hline
$g$&Gravity constant&$\scriptstyle[\frac{\text{m}}{\text{s}^2}]$\\
\hline
$h(x)$&Pipe elevation from $x=0$ to $x=X$&$\scriptstyle[\text{m}]$\\
\hline
$P_i$& Pipe $i$\\
\hline
$R_s$&Specific gas constant&$\scriptstyle[\frac{\text{m}^2}{\text{s}^2\text{K}}]$\\
\hline
$T_0$&Constant temperature&$\scriptstyle[\text{K}]$\\
\hline
$V$&Volume&$\scriptstyle[\text{m}^3]$\\
\hline
$X$&Pipe length&$\scriptstyle[\text{m}]$\\
\hline
$z_0$&Constant compressibility factor&$\scriptstyle[1]$\\
\hline
$\lambda$&Friction factor&$\scriptstyle[1]$\\
\hline
\end{tabular}
\end{center}
\caption{Definitions of variables and their SI-units.}
\label{tab:parameters}
\end{table}
\section{Single pipe}

\begin{tcolorbox}[title=State space model single pipe model]
Under the assumptions below the one-dimensional pipe flow can be described by the following state space model:
\begin{align*}
\dot x 
&= \begin{bmatrix}
0 & -\DpCoeff{}\\
\DqCoeffpr{} & \DqCoeffql{}
\end{bmatrix}x+\begin{bmatrix}
0 & \DpCoeff{}\\
\DqCoeffpl{} & 0
\end{bmatrix}u,\\
y
&=x,
\end{align*}
with $x=\begin{bmatrix}
p_{r}& q_{\ell}
\end{bmatrix}^\top$ as the state and output vector, and $u=\begin{bmatrix}
  p_{\ell} & q_{r}
\end{bmatrix}^\top$ as the input vector.
The coefficients are
\begin{subequations}\label{eq:pipe_coeff}
\begin{align}
\DpCoeff{}&=-\frac{R_sT_0z_0}{AX},\quad \DqCoeffpr{}=-\frac{A}{X},\\
 \DqCoeffpl{}&=\frac{A}{X}+\frac{\lambda R_sT_0z_0}{2DA}\frac{ \bar q\vert  \bar q\vert}{ \bar p_{\ell}^2}-\frac{Agh}{R_sT_0z_0X},\\
\DqCoeffql{}&=-\frac{\lambda R_sT_0z_0}{DA} \frac{|\bar q|}{ \bar p_{\ell}}.
\end{align}
\end{subequations}
\end{tcolorbox}

\subsection{Assumptions}\label{subsec:pipe_ass}
\begin{ass}\label{ass:isoth}
The change in temperature along the pipe is negligible.
\end{ass}
\begin{ass}\label{ass:pipeFlow} For the one-dimensional pipe flow, 
\begin{enumerate}[(i)]
\item the cross-sectional area of each pipe segment is constant;
\item at each point in $x$-dimension averaged velocities suffice;
\item friction along the pipe can be approximated by the Darcy-Weisbach equation, see e.g. \cite{rennels2012pipe};
\item the compressibility factor is constant along the pipe;
\item capillary, magnetic and electrical forces on the fluid are negligible;
\item the gas velocity is much smaller than the speed of sound.
\end{enumerate}
\end{ass}
\subsection{Derivation}
Under the assumptions above Benner \emph{et al.} \cite{benner2019} derive a nonlinear model for an isothermal one-dimensional pipe flow. Towards a discretization and linearization of the related nonlinear dynamics, let the boundary conditions $p_\ell$ and $q_r$ be given, whereas 
$p_r$ and $q$ are to be determined through the model.
Spatial discretization and linearization of \cite[Eq. 3.2]{benner2019} yields
\begin{subequations}\label{eq:linODEs}
\begin{align}
\dot p_r&=\DpCoeff{}(q_r-q_l)\label{eq:dprdt_iso}\\
\dot q_\ell&=
\DqCoeffpr{} p_r+\DqCoeffpl{} p_\ell+\DqCoeffql{} q_\ell,\label{eq:dqldt_iso}
\end{align}
\end{subequations}
or equivalently,
\begin{subequations}
\begin{align*}
\dot x_{t} &= \begin{bmatrix}
0 & -\DpCoeff{}\\
\DqCoeffpr{} & \DqCoeffql{}
\end{bmatrix}x_{t}+\begin{bmatrix}
0 & \DpCoeff{}\\
\DqCoeffpl{} & 0
\end{bmatrix}u_{t},
\end{align*}
\end{subequations}
with terms as described above. 
\subsection{Conservation of mass}
The DC gain from $u$ to $x$ can be readily extracted, 
\begin{align*}
-A^{-1}B=-\frac{1}{\DpCoeff{}\DqCoeffpr{}}\begin{bmatrix}
\DpCoeff{}\DqCoeffpl{} & \DpCoeff{}\DqCoeffql{}\\
0 & -\DpCoeff{} \DqCoeffpr{}
\end{bmatrix}=\begin{bmatrix}
-\frac{\DqCoeffpl{}}{\DqCoeffpr{}} & -\frac{\DqCoeffql{}}{\DqCoeffpr{}}\\
0 & 1
\end{bmatrix},
\end{align*}
which shows that $T_{qp}(0)=0$ and $\one T_{qq}(0)=1$, where
\begin{align*}
X(s)&=T(s)U(s)=\begin{bmatrix}
T_{pp}(s)&T_{pq}(s)\\T_{qp}(s)&T_{qq}(s)
\end{bmatrix}U(s).
\end{align*}
By definition \cite{sven_bob_control} this is conservation of mass.

\section{Branch}\label{sec:branch}
\begin{figure}[ht]
    \centering
   \resizebox{.4\columnwidth}{!}{
  \includegraphics{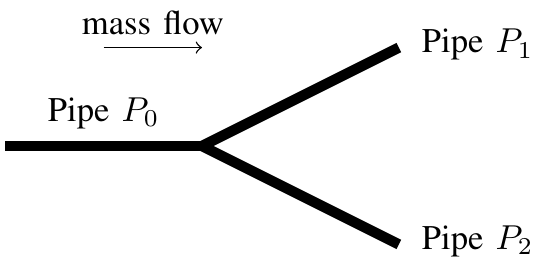}}
 \caption{Branch}
    \label{fig:branch_sketch}
    \end{figure}

Consider a branching pipe geometry as shown in Figure \ref{fig:branch_sketch}.
    
\begin{tcolorbox}[title=State space model simple branch]
Under the assumptions from Section \ref{subsec:pipe_ass} and matching conditions at the junction
\begin{align*}
p_{0,r}&=p_{1,\ell}=p_{2,\ell},\\
q_{0,r}&=q_{1,\ell}+q_{2,\ell},
\end{align*}
the state space realization of the one-dimensional pipe flow through the branch depicted above is as follows:

\begin{subequations}\label{eq:ss_branch}
\begin{align*}
\dot x&=
\begin{multlined}[t]
\begin{bmatrix}0&0&0&
-\DpCoeff{0}& \DpCoeff{0} & \DpCoeff{0}\\
0&0&0&0 & -\DpCoeff{1} & 0\\
0&0&0&0 & 0 & -\DpCoeff{2}\\
\DqCoeffpr{0} & 0 & 0&\DqCoeffql{0} &0&0\\
\DqCoeffpl{1} & \DqCoeffpr{1} & 0&0&\DqCoeffql{1}&0 \\
\DqCoeffpl{2} & 0 & \DqCoeffpr{2}&0&0&\DqCoeffql{2} 
\end{bmatrix}x
+
\begin{bmatrix}
0 & 0 & 0\\
0 & \DpCoeff{1} & 0\\
0 & 0 & \DpCoeff{2}\\
\DqCoeffpl{0} & 0 & 0\\
0 & 0 & 0\\
0 & 0 & 0
\end{bmatrix}u
\end{multlined}\\
y&=\begin{bmatrix}
0 & 1 & 0 & 0 & 0&0\\
0 & 0 & 1 & 0 & 0&0\\
0 & 0 & 0 & 1& 0&0
\end{bmatrix}x,
\end{align*}
\end{subequations}
with state $x=\begin{bmatrix}
p_{0,r} & p_{1,r} & p_{2,r} & q_{0.\ell}&q_{1,\ell}&q_{2,\ell}
\end{bmatrix}^\top$, input $u=\begin{bmatrix}
p_{0,\ell} & q_{1,r} & q_{2,r}
\end{bmatrix}^\top$ and output $y=\begin{bmatrix}
p_{1,r}&p_{2,r}&q_{0,\ell}
\end{bmatrix}^\top.$
\end{tcolorbox}
\subsection{Derivation}    
 Building on \eqref{eq:linODEs}, consider Assumption \ref{ass:pipeFlow} and suppose that additional losses at the intersection are accommodated  by appropriately adjusting the friction factor for all pipes.
Continuity at the boundary of pressure and mass flow means
\begin{subequations}\label{eq:constraints_branch}
\begin{align*}
p_{0,r}&=p_{1,\ell}=p_{2,\ell},\\
q_{0,r}&=q_{1,\ell}+q_{2,\ell},
\end{align*}
\end{subequations}
which sets the input signal of the single pipe model of pipe $P_0$ as the sum of two other state variables. The related matrices for a branch model immediately follow. The extension to a split into more than two pipes is direct and hence for brevity not presented.
\subsection{Conservation of mass}
The state-space realization above satisfies conservation of mass, which follows from the first row of the steady state equation $\zero=Ax+Bu$ and conservation of mass of the single pipe model.

\section{Joint}
We present models first for the case of two joining pipes and then consider the more complex case of multiple pipes joining.
\subsection{Two joining pipes}\label{subsec:joint}
\begin{figure}[ht!]
    \centering
    \resizebox{.4\columnwidth}{!}{
   \includegraphics{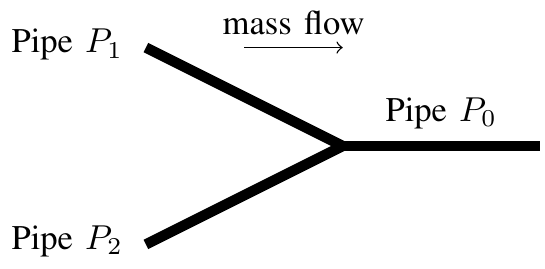}}
 \caption{Joint of 2 pipes merging into one}
    \label{fig:2joiningPipes_sketch}
\end{figure}
\begin{tcolorbox}[title=Simple joint model]
Under the assumptions in Section \ref{subsec:pipe_ass} and boundary conditions
\begin{align*}
p_{1,r}=p_{2,r}=p_{0,\ell},\\
q_{1,r}+q_{2,r}=q_{0,\ell},
\end{align*}
two pipes merging into a single pipe are described by the following state space model:
{\small
\begin{align*}
\dot{\begin{bmatrix}
p_{0,r}\\q_{0,\ell}\\p_{1,r}\\q_{1,\ell}\\q_{2,\ell}
\end{bmatrix}}&=
\begin{multlined}[t]
\begin{bmatrix}
0&-\DpCoeff{0}&0&0&0\\
\DqCoeffpr{0}&\DqCoeffql{0}& \DqCoeffpl{0}&0&0\\
0&\frac{\DpCoeff{1}\DpCoeff{2}}{\DpCoeff{1}+\DpCoeff{2}}&0 & \frac{\DpCoeff{1}^2}{\DpCoeff{1}+\DpCoeff{2}}-\DpCoeff{1}&-\frac{\DpCoeff{1}\DpCoeff{2}}{\DpCoeff{1}+\DpCoeff{2}}\\
0&0&\DqCoeffpr{1}&\DqCoeffql{1}&0\\
0 & 0 & \DqCoeffpr{2} & 0 & \DqCoeffql{2}
\end{bmatrix}
\begin{bmatrix}
p_{0,r}\\q_{0,\ell}\\p_{1,r}\\q_{1,\ell}\\q_{2,\ell}
\end{bmatrix}\\
+
\begin{bmatrix}
0 & 0 & \DpCoeff{0}\\
0 & 0 & 0\\
0 & 0 & 0\\
\DqCoeffpl{1} & 0 & 0\\
0 & \DqCoeffpl{2} & 0
\end{bmatrix}
\begin{bmatrix}
p_{1,\ell}\\p_{2,\ell}\\q_{0,r}
\end{bmatrix},
\end{multlined}\\
\begin{bmatrix}
p_{0,r}\\q_{1,\ell}\\q_{2,\ell}
\end{bmatrix}&=\begin{bmatrix}
1 & 0 & 0 & 0 & 0\\
0 & 0 & 0 & 1 & 0\\
0 & 0 & 0 & 0 & 1\\
\end{bmatrix}
\begin{bmatrix}
p_{0,r}\\q_{0,\ell}\\p_{1,r}\\q_{1,\ell}\\q_{2,\ell}
\end{bmatrix}.
\end{align*}
}
\end{tcolorbox}
\subsubsection{Derivation}
Building on \eqref{eq:linODEs}, each joining pipe can be described by
\begin{subequations}\label{eq:njoint_ode}
\begin{align}
\dot p_{j, r}&=\DpCoeff{j} (q_{j,r}-q_{j,\ell})\label{eq:pdot},\\
\dot q_{j,\ell}&=\DqCoeffpr{j} p_{j,r}+\DqCoeffpl{j} p_{j,\ell}+\DqCoeffql{j} q_{j,\ell}\label{eq:qdot},
\end{align}
\end{subequations}
where the subscript $j$ corresponds to pipe $P_j$, with coefficients as defined above and $j\in\{0,1,\dots,n\}$. 
Every $p_{j,r}$ and $q_{j,\ell}$ are state variables and every $p_{j,\ell}$ and $q_{j,r}$ are input variables related to pipe $P_j$. 

Consider Assumption \ref{ass:pipeFlow} and algebraic constraints
\begin{subequations}
\begin{align}
p_{1,r}=p_{2,r}=p_{0,\ell},\label{eq:ACpn=2}\\
q_{1,r}+q_{2,r}=q_{0,\ell},\label{eq:ACqn=2}
\end{align}
\end{subequations}
which express continuity of the pressure and mass flow. Assume further that additional losses at the intersection are accommodated  by appropriately adjusting the friction factor for all pipes.

Note that \eqref{eq:ACqn=2} cannot directly be incorporated into the state space formulation since it is not immediate how the signal $q_{0,\ell}$ is split between the flow of pipes 1 and 2. To eliminate variables $q_{1,r},q_{r,2}$, consider \eqref{eq:ACpn=2}, take the derivative and use \eqref{eq:pdot} to obtain
\begin{subequations}\label{eq:derived_aeq_n2}
\begin{align}
(\DpCoeff{1}+\DpCoeff{2})q_{1,r}&=\DpCoeff{1}q_{1,\ell}+\DpCoeff{2}(q_{0,\ell}-q_{2,\ell}),\label{eq:q1r_n2}\\
(\DpCoeff{1}+\DpCoeff{2})q_{2,r}&=\DpCoeff{2}q_{2,\ell}+\DpCoeff{1}(q_{0,\ell}-q_{1, \ell})\label{eq:q2r_n2},
\end{align}
\end{subequations}
which yields the state space description above.
\subsubsection{Conservation of mass}
The state-space realization above satisfies conservation of mass, which follows from the third row of the steady state equation $\zero=Ax+Bu$ and conservation of mass of the single pipe model.

\subsection{Multiple joining pipes}
\begin{figure}[ht!]
    \centering
    \resizebox{.4\columnwidth}{!}{
   \includegraphics{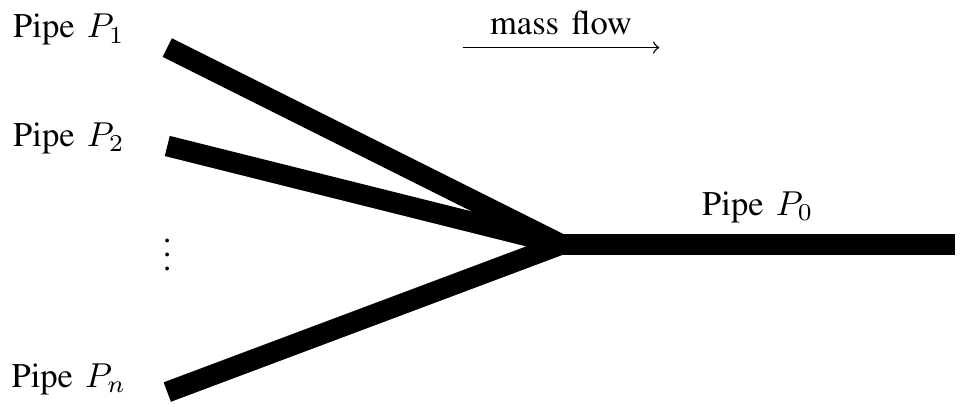}}
 \caption{Joint of $n$ pipes merging into one}
    \label{fig:xjoiningPipes_sketch}
\end{figure}
\begin{tcolorbox}[title=Joint model]
Consider the assumptions in Section \ref{subsec:pipe_ass} and matching conditions at the junction
\begin{align*}
p_{1,r}=p_{2,r}=\dots=p_{n+1,r}=p_{0,\ell},\\
q_{1,r}+q_{2,3}+\dots+q_{r,n}+q_{r,n+1}=q_{0,\ell}.
\end{align*}
 A joint of $n$ pipes with dynamics \eqref{eq:njoint_ode} can be described by the state-space model,
\begin{align*}
\dot{x}&=\begin{bmatrix}
\zero_{2,2} & A_{12}\\
A_{21} & A_{22}
\end{bmatrix} x+\begin{bmatrix}
B_1\\
B_2
\end{bmatrix} u,\\
y&=\begin{bmatrix}
\begin{matrix}
1 & 0 & 0 & 0\\
0 & 0 & 0 & 1
\end{matrix} & \zero_{2,n-1}\\\zero_{n-1,4}&\eye_{n-1}
\end{bmatrix} x,
\end{align*}
where
\begin{align*}
u&=
\begin{bmatrix}
p_{1,\ell}&p_{2,\ell}&\dots&p_{n,\ell}&q_{0,r}
\end{bmatrix}^\top\in\R^{n+1},\\
x&=
\begin{bmatrix}
p_{0,r}&p_{1,r}&q_{0,\ell}&q_{1,\ell}&q_{2,\ell}&\dots&q_{n,\ell}
\end{bmatrix}^\top\in\R^{n+3}, \\
y&=
\begin{bmatrix}
p_{0,r}&q_{1,\ell}&q_{2,\ell}&\dots&q_{n,\ell}
\end{bmatrix}^\top\in\R^{n+1},
\end{align*}
\begin{align*}
A_{12}&=
\begin{bmatrix}
-\DpCoeff{0} & \zero_{n}\\
a & -a\one_{n}
\end{bmatrix},\quad
A_{21}=
\begin{bmatrix}
\DqCoeffpr{0} & 0 & 0 & \dots & 0\\
\DqCoeffpl{0} &  \DqCoeffpr{1} &  \DqCoeffpr{2}& \dots & \DqCoeffpr{n}
\end{bmatrix}^\top,\quad
A_{22}=\diag(\DqCoeffql{0} ,\dots,\DqCoeffql{n}) \\
B_1&=
\begin{bmatrix}
\zero_n& \DpCoeff{0} \\
\zero_n& 0
\end{bmatrix},\quad 
B_2=\begin{bmatrix}
\zero_{n+1}\\
\begin{matrix}
\begin{matrix}
\DqCoeffpl{1}&&\\
&\ddots&\\
&&\DqCoeffpl{n}
\end{matrix}&\zero_n^\top
\end{matrix}
\end{bmatrix},
\end{align*}
with
\begin{align*}
a&=\DpCoeff{1}\left(\sum_{j=1}^n\prod_{\substack{i=1\\i\neq j}}^n\DpCoeff{i}\right)^{-1}\prod_{\substack{i=2}}^n \DpCoeff{i}.
\end{align*}
\end{tcolorbox}
\subsubsection{Derivation}
We wish to extend the model for $n=2$ joining pipes to any finite number of joining pipes. In particular, first we would like to obtain an expression for $q_{j,r}$ only depending on state variables and sources, as in \eqref{eq:derived_aeq_n2}. 
\begin{defin}
 We say \textup{sources} for inputs into the joint model that are not related to states of any other pipes via algebraic constraints. We say \textup{internal variables} for input variables that are not sources. 
\end{defin} 
We derive the model by induction. Consider the algebraic constraints for $n=3$,
\begin{align*}
p_{1,r}=p_{2,r}=p_{3,r}=p_{0,\ell},\\
q_{1,r}+q_{2,r}+q_{3,r}=q_{0,\ell}.
\end{align*}
Then,
\begin{align*}
\dot p_{1,r}&=\dot p_{2,r},\\
(\DpCoeff{1}+\DpCoeff{2})q_{1,r}&=\DpCoeff{1}q_{1,\ell}+\DpCoeff{2}(q_{0,\ell}-q_{2,\ell}-q_{3,r}),\\[1em]
\dot p_{3,r}&=\dot p_{2,r},\\
(\DpCoeff{2}+\DpCoeff{3})q_{3,r}&=\DpCoeff{3}q_{3,\ell}+\DpCoeff{2}(q_{0,\ell}-q_{2,\ell}-q_{1,r})\\
&=\begin{multlined}[t]
\DpCoeff{3}q_{3,\ell}+\DpCoeff{2}\Big(q_{0,\ell}-q_{2,\ell}-(\DpCoeff{1}+\DpCoeff{2})^{-1}
[\DpCoeff{1}q_{1,\ell}+\DpCoeff{2}(q_{0,\ell}-q_{2,\ell}-q_{3,r})]\Big).\end{multlined}
\end{align*}
so that
\begin{align}
\begin{multlined}[t]
(\DpCoeff{1}\DpCoeff{2}+\DpCoeff{1}\DpCoeff{3}+\DpCoeff{2}\DpCoeff{3})q_{3,r}=\DpCoeff{1}\DpCoeff{2}(q_{0,\ell}-q_{1,\ell}-q_{2,\ell})
+(\DpCoeff{1}\DpCoeff{3}+\DpCoeff{2}\DpCoeff{3})q_{3,\ell}.\end{multlined}\label{eq:q3r_n3}
\end{align}
Similarly,
\begin{align}
\begin{multlined}[t]
(\DpCoeff{1}\DpCoeff{2}+\DpCoeff{1}\DpCoeff{3}+\DpCoeff{2}\DpCoeff{3})q_{1,r}=\DpCoeff{2}\DpCoeff{3}(q_{0,\ell}-q_{2,\ell}-q_{3,\ell})
+(\DpCoeff{1}\DpCoeff{2}+\DpCoeff{1}\DpCoeff{3})q_{1,\ell},\end{multlined}\label{eq:q1r_n3}\\
\begin{multlined}[t]
(\DpCoeff{1}\DpCoeff{2}+\DpCoeff{1}\DpCoeff{3}+\DpCoeff{2}\DpCoeff{3})q_{2,r}=\DpCoeff{1}\DpCoeff{3}(q_{0,\ell}-q_{1,\ell}-q_{3,\ell})
+(\DpCoeff{1}\DpCoeff{2}+\DpCoeff{2}\DpCoeff{3})q_{2,\ell}.\end{multlined}\label{eq:q2r_n3}
\end{align}
Comparing equations \eqref{eq:q3r_n3}--\eqref{eq:q2r_n3} with \eqref{eq:derived_aeq_n2} we recognize a certain pattern, which we explore next. Facilitated by observations above we derive a general statement.
\begin{prop}\label{prop:qkr}
Consider $n$ joining pipes with individual dynamics \eqref{eq:pdot} - \eqref{eq:qdot}. Let $q_{j,\ell},\, j\in\{0,1,\dots,n\},$ be state variables. The internal variables $q_{k,r},\,k\in\{1,2,\dots,n\},$ can be described by state variables,
\begin{align}
\begin{multlined}[t]
\left(\sum_{j=1}^n\prod_{\substack{i=1\\i\neq j}}^n\DpCoeff{i}\right)q_{k,r}=\prod_{\substack{i=1\\i\neq k}}^n \DpCoeff{i}\left(q_{0,\ell}-\sum_{\substack{i=1\\i\neq k}}^n q_{i,\ell}\right)
+\left(\sum_{j=1}^n\prod_{\substack{i=1\\i\neq j}}^n\DpCoeff{i}-\prod_{\substack{i=1\\i\neq k}}^n\DpCoeff{i}\right)q_{k,\ell}\end{multlined}\label{eq:qkr}
\end{align}
\end{prop}
\begin{proof}
We prove the theorem by induction. We observe that \eqref{eq:qkr} holds for $n\in\{2,3\}$. Hence, assume it holds for some $n$ and for $n+1$ note that interconnections dictate the algebraic constraints
\begin{subequations}
\begin{align}
p_{1,r}=p_{2,r}=\dots=&p_{n+1,r}=p_{0,\ell},\label{eq:ac_press_n}\\
q_{1,r}+q_{2,3}+\dots+q_{r,n}+q_{r,n+1}&\doteq q^n_{0,\ell}+q_{r,n+1}=q_{0,\ell}.
\end{align}
\end{subequations}
Hence, by the algebraic constraints on the pressure and the pipe dynamics, 
\begin{align*}
q_{n+1,r}&=\frac{1}{\DpCoeff{n+1}}(\dot p_{k,r}-\dot p_{n+1,r})+q_{n+1,r}\\
&=\frac{1}{\DpCoeff{n+1}}\left(\DpCoeff{k}(q_{k,r}-q_{k,\ell})-\dot p_{n+1,r}\right)+q_{n+1,\ell}+\frac{\dot p_{n+1,r}}{\DpCoeff{n+1}}\\
&=\frac{1}{\DpCoeff{n+1}}\DpCoeff{k}(q_{k,r}-q_{k,\ell})+q_{n+1,\ell},
\end{align*}
where without loss of generality $k\in\{1,2\dots,n-1\}$.
Then \eqref{eq:qkr} yields
{\small
\begin{align*}
\left(\sum_{j=1}^n\prod_{\substack{i=1\\i\neq j}}^n\DpCoeff{i}\right)q_{k,r}&=\begin{multlined}[t]
\prod_{\substack{i=1\\i\neq k}}^n \DpCoeff{i}\left(q^n_{0,\ell}-\sum_{\substack{i=1\\i\neq k}}^n q_{i,\ell}\right)
+\left(\sum_{j=1}^n\prod_{\substack{i=1\\i\neq j}}^n\DpCoeff{i}-\prod_{\substack{i=1\\i\neq k}}^n\DpCoeff{i}\right)q_{k,\ell}
\end{multlined}\\
&=\begin{multlined}[t]
\prod_{\substack{i=1\\i\neq k}}^n \DpCoeff{i}\left(q_{0,\ell}-q_{n+1,r}-\sum_{\substack{i=1\\i\neq k}}^n q_{i,\ell}\right)
+\left(\sum_{j=1}^n\prod_{\substack{i=1\\i\neq j}}^n\DpCoeff{i}-\prod_{\substack{i=1\\i\neq k}}^n\DpCoeff{i}\right)q_{k,\ell}
\end{multlined}\\
&=\begin{multlined}[t]
-\prod_{\substack{i=1}}^n \DpCoeff{i}\frac{1}{\DpCoeff{n+1}}(q_{k,r}-q_{k,\ell})
+ \prod_{\substack{i=1\\i\neq k}}^n \DpCoeff{i}\left(q_{0,\ell}-q_{n+1,\ell}-\sum_{\substack{i=1\\i\neq k}}^n q_{i,\ell}\right)
+\left(\sum_{j=1}^n\prod_{\substack{i=1\\i\neq j}}^n\DpCoeff{i}-\prod_{\substack{i=1\\i\neq k}}^n\DpCoeff{i}\right)q_{k,\ell}
\end{multlined}\\
&=\begin{multlined}[t]
-\frac{1}{\DpCoeff{n+1}}\prod_{\substack{i=1}}^n \DpCoeff{i}q_{k,r}
+ \prod_{\substack{i=1\\i\neq k}}^n \DpCoeff{i}\left(q_{0,\ell}-\sum_{\substack{i=1\\i\neq k}}^{n+1} q_{i,\ell}\right)
+\left(\sum_{j=1}^n\prod_{\substack{i=1\\i\neq j}}^n\DpCoeff{i}-\prod_{\substack{i=1\\i\neq k}}^n\DpCoeff{i}\right.
\left.+\frac{1}{\DpCoeff{n+1}}\prod_{\substack{i=1}}^n \DpCoeff{i}\right)q_{k,\ell}\end{multlined}.
\end{align*}
}
Move the first term of the right-hand side to the left, multiply the equation by $\DpCoeff{n+1}$ and observe that the left-hand side is
\begin{align*}
\sum_{j=1}^n\prod_{\substack{i=1\\i\neq j}}^n\DpCoeff{i}\DpCoeff{n+1}+\prod_{\substack{i=1}}^n \DpCoeff{i}=\sum_{j=1}^n\prod_{\substack{i=1\\i\neq j}}^{n+1}\DpCoeff{i}+\prod_{\substack{i=1}}^n \DpCoeff{i}=\sum_{j=1}^{n+1}\prod_{\substack{i=1\\i\neq j}}^{n+1}\DpCoeff{i}.
\end{align*}
This identity can also be used for the last term on the right-hand side. Then,
\begin{align*}
\left(\sum_{j=1}^{n+1}\prod_{\substack{i=1\\i\neq j}}^{n+1}\DpCoeff{i}\right)q_{k,r}&=
\begin{multlined}[t]
\prod_{\substack{i=1\\i\neq k}}^{n+1} \DpCoeff{i}\left(q_{0,\ell}-\sum_{\substack{i=1\\i\neq k}}^{n+1} q_{i,\ell}\right)
+ \left(\sum_{j=1}^{n+1}\prod_{\substack{i=1\\i\neq j}}^{n+1}\DpCoeff{i}-\prod_{\substack{i=1\\i\neq k}}^{n+1}\DpCoeff{i}\right)q_{k,\ell},
\end{multlined}
\end{align*}
which is equivalent to our induction hypothesis and thus concludes the proof.
\end{proof}
The state space realization above is a direct consequence of Proposition \ref{prop:qkr}, the pipe dynamics in \eqref{eq:njoint_ode}, the algebraic constraint of equal pressure at the intersection in \eqref{eq:ac_press_n}, and the observation that for the ODE of $p_{1,r}$,
\begin{align*}
%
\begin{multlined}[t]
\DpCoeff{1}\left(\sum_{j=1}^n\prod_{\substack{i=1\\i\neq j}}^n\DpCoeff{i}\right)^{-1}\left(\sum_{j=1}^n\prod_{\substack{i=1\\i\neq j}}^n\DpCoeff{i}-\prod_{\substack{i=2}}^n\DpCoeff{i}\right)-\DpCoeff{1}
%
=-\DpCoeff{1}\left(\sum_{j=1}^n\prod_{\substack{i=1\\i\neq j}}^n\DpCoeff{i}\right)^{-1}\prod_{\substack{i=2}}^n \DpCoeff{i}.\end{multlined}
\end{align*} 

\subsubsection{Conservation of mass}
As for $n=2$ joining pipes, the state-space realization above satisfies conservation of mass, which follows from the third row of the steady state equation $\zero=Ax+Bu$ and conservation of mass of the single pipe model.
\section{Star junction}
\begin{figure}[ht!]
    \centering
    \resizebox{.4\columnwidth}{!}{
   \includegraphics{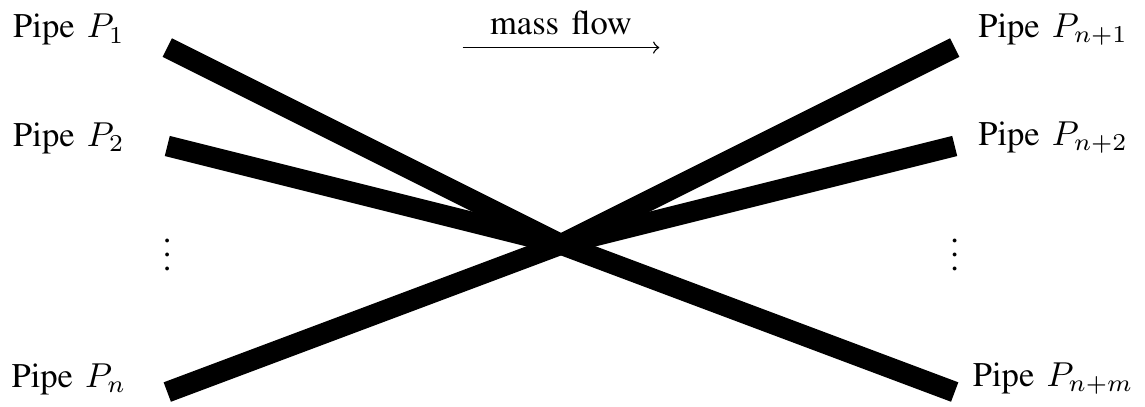}}
 \caption{Star junction of $n+m$ pipes}
    \label{fig:star_junction_sketch}
\end{figure}

Consider the pipe geometry illustrated in Figure \ref{fig:star_junction_sketch}, combining the branch and joint model to so called star junction.
\begin{tcolorbox}[title=Star junction model]
Consider the assumptions in Section \ref{subsec:pipe_ass} and matching conditions at the junction
\begin{align*}
p_{1,r}=\dots=p_{n,r}=p_{n+1,\ell}=\dots=p_{n+m,\ell},\\
q_{1,r}+\dots+q_{r,n}=q_{n+1,\ell}+\dots+q_{n+m,\ell}.
\end{align*}
A star junction of $n$ pipes connecting to $m$ pipes with individual dynamics \eqref{eq:pdot} - \eqref{eq:qdot} can be represented by the state-space model,
\begin{align*}
\dot x_1&=A_1x_1+B_1u,\\
\dot x_2&=A_1x_2+B_2u,\\
\dot x_3&=A_1x_3+B_3u,\\
\dot x_4&=A_1x_4+B_4u,\\
y&=C(x_1+x_2+x_3+x_4),
\end{align*}
where
{\small
\begin{align*}
x_1&=x_2=x_3=x_4\\
&=\begin{multlined}[t]
[\begin{matrix}
p_{1,r}&p_{n+1,r}&p_{n+2,r}&\dots&p_{n+m,r}&q_{1,\ell}\end{matrix}
\begin{matrix}
q_{2,\ell}&\dots&q_{n+m,\ell}
\end{matrix}]^\top\in\R^{n+2m+1},
\end{multlined}\\
u&=\begin{multlined}[t]
[\begin{matrix}
p_{1,\ell}&p_{2,\ell}&\dots&p_{n,\ell}&q_{n+1,r}&q_{n+2,r}\end{matrix}
\begin{matrix}
\dots&q_{n+m,r}
\end{matrix}]^\top\in\R^{n+m},
\end{multlined} \\
y&=
\begin{multlined}[t]
[\begin{matrix}
p_{n+1,r}&p_{n+2,r}&\dots&p_{n+m,r}&q_{1,\ell}&q_{2,\ell}\end{matrix}
\begin{matrix}
\dots&q_{n,\ell}
\end{matrix}]^\top\in\R^{n+m},\end{multlined}
\end{align*}
\begin{align*}
A_1&=
\begin{bmatrix}
\begin{array}{c}
\begin{matrix}
\mathbf{0}_{m+1} & -a & -a\mathbf{1}_{n-1} & a\mathbf{1}_{m}
\end{matrix}\\
\hline
\mathbf{0}_{n+2m,n+2m+1}
\end{array}
\end{bmatrix},
\quad
B_1=\mathbf{0}_{n+2m+1,n+m},
\end{align*}
\begin{align*}
A_2 &= \begin{bmatrix}
\begin{array}{c|c}
\mathbf{0}_{n+2m+1,n+m+1} &
\begin{array}{c}
\begin{matrix}
\mathbf{0}_{m}\\
\hline
\begin{matrix}
-\DpCoeff{n+1}&&\mathbf{0}\\
&\ddots&\\
\mathbf{0}&&-\DpCoeff{n+m}
\end{matrix}
\end{matrix}\\
\hline
\mathbf{0}_{n+m}
\end{array} 
\end{array}
\end{bmatrix},\quad
B_2=
\begin{bmatrix}
\begin{array}{c}
\mathbf{0}_{n+m}\\
\hline
\begin{array}{c|c}
\mathbf{0}_{m,n}
&
\begin{matrix}
\DpCoeff{n+1}&&\mathbf{0}\\
&\ddots&\\
\mathbf{0}&&\DpCoeff{n+m}
\end{matrix}
\end{array}\\
\hline
\mathbf{0}_{n+m,n+m}
\end{array}
\end{bmatrix},
\end{align*}
\begin{align*}
A_3&=
\begin{bmatrix}
\begin{array}{c}
\mathbf{0}_{m+1,n+2m+1}\\
\hline
\begin{array}{c|c|c|c}
\begin{matrix}
\DqCoeffpr{1}\\
\vdots\\
\DqCoeffpr{n}
\end{matrix}
&
\mathbf{0}_{n,m}
&
\begin{matrix}
\DqCoeffql{1}&&\mathbf{0}\\
&\ddots&\\
\mathbf{0} &&\DqCoeffql{n}
\end{matrix}
&
\mathbf{0}_{n,m}
\end{array}\\
\hline
\mathbf{0}_{m,n+2m+1}
\end{array}
\end{bmatrix}, \quad
B_3=
\begin{bmatrix}
\begin{array}{c}
\mathbf{0}_{m+1,n+m}\\
\hline
\begin{array}{c|c}
\begin{matrix}
\DqCoeffpl{1}&&\mathbf{0}\\
&\ddots&\\
\mathbf{0}&&\DqCoeffpl{n}
\end{matrix}
&
\mathbf{0}_{n,m}
\end{array}\\
\hline
\mathbf{0}_{m,n+m}
\end{array}
\end{bmatrix},
\end{align*}
\begin{align*}
A_4&=\begin{multlined}[t]
\left[
\begin{matrix}
\begin{array}{c}
\mathbf{0}_{n+m+1,n+2m+1}\\
\hline
\begin{array}{c|c|c|}
\begin{matrix}
\DqCoeffpl{n+1}\\
\vdots\\
\DqCoeffpl{n+m}
\end{matrix}
&
\begin{matrix}
\DqCoeffpr{n+1}&&\mathbf{0}\\
&\ddots&\\
\mathbf{0} &&\DqCoeffpr{n+m}
\end{matrix}
&
\mathbf{0}_{m,n}
\end{array}
\end{array}
\end{matrix}\right.
\left.
\begin{matrix}
\begin{array}{c}\\
\hline
\begin{matrix}
\DqCoeffql{n+1}&&\mathbf{0}\\
&\ddots&\\
\mathbf{0} &&\DqCoeffql{n+m}
\end{matrix}
\end{array}
\end{matrix}\right]
\end{multlined},\quad
B_4=\mathbf{0}_{n+2m+1,n+m},
\end{align*}
\begin{align*}
C&=\begin{bmatrix}
\mathbf{0}_{n+m} & I_{n+m} & \mathbf{0}_{n+m,m}
\end{bmatrix},
\end{align*}
}
\end{tcolorbox}

\subsubsection{Derivation}
Proposition \ref{prop:qkr} also facilitates a statement about a general star junction, which consists of $n$ pipes connecting to $m$ pipes, as illustrated in Figure \ref{fig:star_junction_sketch}. Towards a state-space realization and akin to the case of a joint, we wish to express internal variables in terms of state variables.
\begin{corollary}\label{col:pkr_qkr}
Consider a star junction with $n$ joining pipes connecting to $m$ branching pipes with individual dynamics \eqref{eq:pdot} - \eqref{eq:qdot}. Let $q_{i,\ell},\, i\in\{1,2,\dots,n+m\}$, and $p_{1,r}$ be state variables. The internal variables $q_{k,r}, \,k\in\{1,2,\dots,n\}$ and $p_{j,\ell},\,j\in\{n+1,n+2,\dots,n+m\},$ can be described by state variables through 
\begin{align}
\left(\sum_{j=1}^n\prod_{\substack{i=1\\i\neq j}}^n\DpCoeff{i}\right)q_{k,r}&=
\begin{multlined}[t]
\prod_{\substack{i=1\\i\neq k}}^n \DpCoeff{i}\left(\sum_{i=n+1}^{n+m}q_{i,\ell}-\sum_{\substack{i=1\\i\neq k}}^n q_{i,\ell}\right)
+\left(\sum_{j=1}^n\prod_{\substack{i=1\\i\neq j}}^n\DpCoeff{i}-\prod_{\substack{i=1\\i\neq k}}^n\DpCoeff{i}\right)q_{k,\ell},\end{multlined}\label{eq:qkr_star_junction}\\
p_{j,\ell}&=p_{1,r}\label{eq:pjl_star}.
\end{align}
\end{corollary}
\begin{proof}
Consider \eqref{eq:qkr} from Proposition \ref{prop:qkr} and note that therein that $q_{0,\ell}=\sum_{j=n+1}^{n+m}q_{j,\ell}$. The result for the first equation follows immediately. The second equation is a direct consequence of the algebraic constraint of all pressures being equal at the intersection.
\end{proof}
The state space realization above follows directly: the system related to $x_1$ describes the solution to $\dot p_{1,r}$ and directly follows from \eqref{eq:qkr_star_junction}. The system corresponding to $x_2$ describes $p_{j,r},\,j\in\{n+1,n+2,\dots,n+m\}$ and is immediately obtained by the dynamics \eqref{eq:pdot}. The systems for $x_3$ and $x_4$ relate to $q_{j,\ell},\,j\in\{1,2,\dots,n+m\}$ and are a consequence of the dynamics \eqref{eq:qdot} and the algebraic constraint \eqref{eq:pjl_star}.
\subsubsection{Conservation of mass}
Conservation of mass can be established from the branch and joint above and is omitted for brevity.

\section{Control valve}\label{sec:valve}
We present two models for control valves: firstly, we use a static gain (less or equal than 1) on the pressure and unity gain on mass flow; secondly, we provide a more complex model with the mass flow varying with in- and outlet pressure and cross-sectional area.
\subsection{Static model}
\begin{tcolorbox}[title=Static valve model]
For the static gain model let the states and inputs be:
\begin{align}
x=\begin{bmatrix}
p_r&q_\ell
\end{bmatrix}^\top,\quad u=\begin{bmatrix}
p_\ell & q_r
\end{bmatrix}^\top,
\end{align}
and define the static relationship
\begin{align*}
x=\begin{bmatrix}
k_v&0\\0 & 1
\end{bmatrix}u,
\end{align*}
where $k_v\in(0,1)$.
\end{tcolorbox}
Conservation of mass is immediate. 
\subsection{Dynamic model}
Here we develop a model based on a static relationship commonly used for orifices combined with a first order low pass filter approximating the dynamics between command and actuation. 

\begin{tcolorbox}[title=Dynamic valve model
\footnotemark] Assume an isentropic expansion and approximate the command-to-actuation relationship by a first order model. Define the state, about and input signals as
\begin{align*}
x=\begin{bmatrix}
A_o
\end{bmatrix},\quad y=\begin{bmatrix}
q_v
\end{bmatrix},\quad u=\begin{bmatrix}
u_v&p_{\ell}&p_{r}
\end{bmatrix}^\top.
\end{align*}
The dynamics are 
\begin{subequations}\label{eq:valv_linmod}
\begin{align}
\dot x&=\begin{bmatrix}
-1/\tau
\end{bmatrix}x+\begin{bmatrix}
K/\tau & 0& 0
\end{bmatrix}u,\\
 y&=\begin{multlined}[t]\begin{bmatrix}
g_o(A_o,\bar p_{\ell},\bar p_{r})/A_o
\end{bmatrix}x
+\begin{bmatrix}
0&\zeta_o(A_o,\bar p_{\ell},\bar p_{r})&\xi_o(A_o,\bar p_{\ell},\bar p_{r})
\end{bmatrix}u,\end{multlined}
\end{align}
\end{subequations}
where $g_0$ is the orifice equation from \eqref{eq:q-orifice}, $\tau$ and $K$ the time constant and gain from transfer function \eqref{eq:lpf_actuator} between the control command and actual actuation, $q_v$ the mass flow through the valve, $u_v\in(0,1)$ the control input, $A_o$ the cross-sectional area of the valve, and $\xi_o$ and $\zeta_o$ the linearization terms from \eqref{eq:orifice_lin_pr} and \eqref{eq:orifice_lin_pl}.
\end{tcolorbox}
\footnotetext{The dynamic valve model has not been fully tested in operational models.}
We note that conservation of mass does not apply to this model, as there are no mass flows as model inputs.

\subsubsection{Derivation}

The static mapping is justified by pipe time constants and the usual sampling time, which both vastly exceed the low time constants of the valve \cite{funk72_valve}.

In \cite[Equ. (3-34)]{mattingly}, assuming an isentropic expansion the mass flow through an orifice is approximated by 
{\small
\begin{align}
q_o&=Cp_{o,\ell} A_o\sqrt{\frac{2}{R_sT_0z_0}\frac{\mu}{\mu-1}\left[\left(\frac{p_{o,r}}{p_{o,\ell}}\right)^{2/\mu}-\left(\frac{p_{o,r}}{p_{o,\ell}}\right)^{(\mu+1)/\mu}\right]},\label{eq:q-orifice}\\
&\doteq g_o(A_o,p_r,p_\ell)
\end{align}
}
where $C\doteq C_d/\sqrt{1-(D_0/D_1)^4}$ corrects for the head loss, with $C_d$ being the discharge coefficient, $\mu$ the ratio of specific heat, $\mu=c_p/c_v$ and $D_0(D_1)$ the orifice (pipe) diameter. 

For the control valve let the cross-sectional area, $A_o$, from \eqref{eq:q-orifice} represent the manipulated variable. To approximate the dynamics of the actuator itself, i.e., the relationship between the control command and actual actuation, consider a first-order low pas filter with transfer function,
\begin{align}
K\frac{1}{\tau s+1},\label{eq:lpf_actuator}
\end{align}
where $\tau>0$ is the time constant, $K=A_{o,max}$ is the gain. The control input is constrained, i.e., $u_v\in(0,1)$, where $0$ is closed and $1$ is open.
Towards a linear model:
\begin{align}
\frac{\partial g_o(A_o,p_\ell,p_r)}{\partial p_r}&\doteq \xi_o(A_o,p_\ell,p_r),\label{eq:orifice_lin_pr}\\
\frac{\partial g_o(A_o,p_\ell,p_r)}{\partial p_\ell}&\doteq \zeta_o(A_o,p_\ell,p_r).\label{eq:orifice_lin_pl}
\end{align}
Combining these equations with \eqref{eq:lpf_actuator} yields the state space model above.

\section{Tank}
We next introduce two tank models: the first model approximates pressure changes while assuming a constant temperature; the second model additionally admits temperature changes. 
\subsection{Isothermal tank}
\begin{tcolorbox}[title=Isothermal tank model]
Assume there are multiple in- and outlets at the tank and the temperature is constant. Under the assumptions below, for the pressure as the single state, $x=p$, and input vector
\begin{align*}
u=\begin{multlined}[t]\left[\begin{matrix}
q_{1,\ell}\dots&q_{n_\ell,\ell}&q_{1,r}&\dots&q_{n_r,r}
\end{matrix}\right],\end{multlined}
\end{align*}
with $n_\ell\in\mathbb{N}$ as the number of inlets and $n_r\in\mathbb{N}$ as the number of outlets, the linear state space model is
\begin{align*}
\dot x=\frac{R_s z_0T_0}{V}\begin{bmatrix}
\one_{n_\ell} & -\one_{n_r}
\end{bmatrix} u,
\end{align*}
with $V$ as the constant volume.
\end{tcolorbox}
\subsubsection{Assumptions}
\begin{ass}\label{ass:tank} Suppose:
\begin{enumerate}[label=\ref{ass:tank}(\roman*)]
\item there is perfect mixing inside the volume;\label{ass:mix}
\item the compressibility factor, $z$, changes negligibly or if necessary can be represented by a time-varying parameter;\label{ass:z}
\item the gas is described by the ideal gas equation; \label{ass:rge}
\end{enumerate}
\end{ass}
\subsubsection{Derivation}
For the pressure inside the constant volume, we have 
\begin{align}
p&=\rho R_s T_0z_0,\nonumber\\
\dot p&=  R_s z_0T_0\dot \rho
 = R_s z_0T_0\frac{q}{V} 
 = \frac{R_s z_0T_0}{V} \left(\sum_{j=1}^{n_\ell}q_{j,\ell} - \sum_{k=1}^{n_r}q_{k,r}\right)\label{eq:p_tank_iso}
\end{align}
where the first equation is the ideal gas equation. The derivative uses Assumption \ref{ass:z} and the last line conservation of mass: change in mass equals mass flow in minus mass flow out. The variables are: $\rho$ density; $\Sigma_jq_{j,\ell}$ ($\Sigma_kq_{k,r}$) sum of mass flowing in (out). The state space model above and conservation of mass follow directly.
\subsection{Non-isothermal tank}
We now drop the isothermal assumption and incorporate temperature dynamics. This also has an effect on the pressure dynamics themselves.

\begin{tcolorbox}[title=Non-isothermal tank model]
Consider Assumptions \ref{ass:tank} and \ref{ass:noniso_tank}. We obtain a nonlinear representation relating the entering mass flows and temperatures, $q_{i,\ell}$ and $\;T_{i,\ell},i=\{1,\dots,n_\ell\}$, and exiting mass flows, $q_{i,r},i=\{1,\dots,n_r\}$, to the pressure and temperature inside the tank, $p$ and $T$: 
\begin{align*}
\dot p &= \frac{R_s z_0\mu}{V} \left(\sum_{j=1}^{n_\ell}q_{j,\ell}T_{j,\ell} - \sum_{k=1}^{n_r}q_{k,r} T\right),\\
\dot T &=\begin{multlined}[t]\frac{R_sTz_0}{pVc_v}\left(\sum_{j=1}^{n_\ell}q_{i,j}\left(c_{p,j}T_j -c_vT\right)- R_s T\sum_{k=1}^{n_r}q_{o,k}\right).\end{multlined}
\end{align*}
Therefore, with $x=\begin{bmatrix}
p & T
\end{bmatrix}^\top$ and
\begin{align*}
u=\begin{multlined}[t]\left[\begin{matrix}
q_{1,\ell}&T_{1,\ell}&\dots&q_{n_\ell,\ell}&T_{n_i,\ell}\end{matrix}\qquad \right.
\left.\begin{matrix}&q_{1,r}&\dots&q_{n_r,r}
\end{matrix}\right],\end{multlined}
\end{align*}
we can use the linearized dynamics for pressure and temperature to generate a linear state space model.
\end{tcolorbox}

\subsubsection{Assumptions}
\begin{ass}\label{ass:noniso_tank} Suppose:
\begin{enumerate}[label=\ref{ass:noniso_tank}(\roman*)]
\item potential and kinetic energy are negligible;\label{ass:energy}
\item the reference states for internal energy and enthalpy  are set to zero at zero degree Kelvin; \label{ass:t0u0}
\item specific heats are constant;\label{ass:cv}
\item the heat flux to the environment is negligible. \label{ass:Qdot_zero}
\end{enumerate}
\end{ass}

\subsubsection{Derivation}

The pressure dynamics are akin to the isothermal case but differ as the temperature is not a constant anymore:
\begin{align}
\dot p &=  R_s z_0\left(\frac{\sum_{j=1}^{n_\ell}q_{j,\ell} - \sum_{k=1}^{n_r}q_{k,r}}{V} T+\rho \dot T\right)\nonumber\\
 &= R_s z_0\left(\frac{\sum_{j=1}^{n_\ell}q_{j,\ell} - \sum_{k=1}^{n_r}q_{k,r}}{V} T+\frac{p}{R_sTz_0} \dot T\right)\label{eq:pdot_w_Tdot}
\end{align}
We next derive an expression for $\dot T$, which we will plug back in.

\paragraph*{Temperature dynamics}
As mentioned in Assumptions~\ref{ass:rge} and \ref{ass:t0u0} we assume ideal gas and select the reference states for the internal energy and enthalpy to be zero at zero degree Kelvin: $T_{\rm ref}=0\implies U_{\rm ref},H_{\rm ref}=0$ (see for example \cite[Ch. 3.6.3]{moranfundamentals}). Then, the derivative of the total energy of a system, $E$, is \cite[Equ. (2.27)]{moranfundamentals}:
\begin{align}
\frac{d}{dt}E&\stackrel{Ass. \ref{ass:energy}}{=}\frac{d}{dt}U\\
&\stackrel{Ass. \,\ref{ass:mix}, \ref{ass:t0u0}}{=}\frac{d}{dt}(m c_vT)\nonumber\\
&\stackrel{Ass. \,\ref{ass:cv}}{=}\left(\sum_{j=1}^{n_\ell}q_{j,\ell} - \sum_{k=1}^{n_r}q_{k,r}\right)c_vT+mc_v\dot T\label{eq:Udot_tot_energy}
\end{align}
with $U$ being the internal energy. Moreover, conservation of energy implies \cite[Equ. (4.9)-(4.12)]{moranfundamentals}
\begin{align}
\frac{d}{dt}U&\stackrel{Ass. \ref{ass:energy}}{=} \dot Q+ \sum_{j=1}^{n_\ell}q_{j,\ell}h_j - \sum_{k=1}^{n_r}q_{k,r}h_k\nonumber\\
&\stackrel{Ass. \ref{ass:mix}}{=}\dot Q +c_{p}\sum_{j=1}^{n_\ell}q_{j,\ell}T_j - c_pT\sum_{k=1}^{n_r}q_{k,r},\label{eq:Udot_conserve}
\end{align}
where $h$ is the specific enthalpy, and $\dot Q$ is the rate of heat transfer to the environment, which is a function of the temperature (deviations) inside and outside the volume. Usually, empirical formulations are used for conduction, convection and radiation (see \cite[Ch. 2]{moranfundamentals}). Here for brevity we assume it is negligible (cf. Assumption \ref{ass:Qdot_zero}).
Combining \eqref{eq:Udot_tot_energy} and \eqref{eq:Udot_conserve} yields
\begin{align}
\dot T&=\begin{multlined}[t]\frac{1}{mc_v}\left[c_vT \left(\sum_{k=1}^{n_r}q_{k,r}-\sum_{j=1}^{n_\ell}q_{j,\ell}\right)+\sum_{j=1}^{n_\ell}q_{j,\ell}c_{p}T_j \right.
\left.- c_pT\sum_{k=1}^{n_r}q_{k,r}\right]\end{multlined}\nonumber\\
&=\begin{multlined}[t]\frac{1}{mc_v}\left[\sum_{j=1}^{n_\ell}q_{j,\ell}\left(c_{p}T_j -c_vT\right)\ - R_s T\sum_{k=1}^{n_r}q_{k,r}\right]\end{multlined}\nonumber\\
&=\begin{multlined}[t]\frac{R_sTz_0}{pVc_v}\left[\sum_{j=1}^{n_\ell}q_{j,\ell}\left(c_{p}T_j -c_vT\right)- R_s T\sum_{k=1}^{n_r}q_{k,r}\right],\end{multlined}\label{eq:Tdot}
\end{align}
which represents the nonlinear temperature dynamics from the model. Note that \eqref{eq:Tdot} also coincides with that in \cite[Ch. 13.4.5]{egeland2002}. Using this equation in \eqref{eq:pdot_w_Tdot} yields the stated nonlinear dynamics for the pressure.

\section{Valve manifold}\label{sec:composite_models}
\begin{figure}[ht!]
	\centering
   \resizebox{.7\columnwidth}{!}
   {\includegraphics{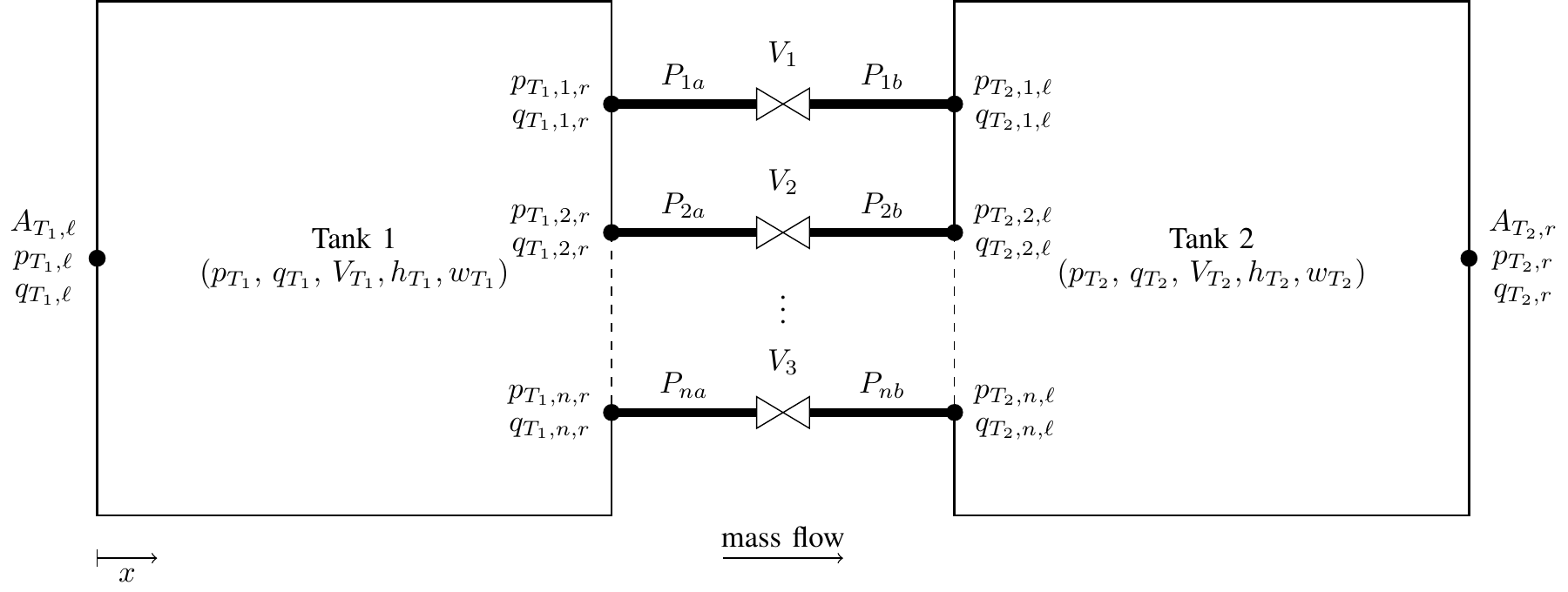}}
 \caption{Manifold}
    \label{fig:manifold}
\end{figure}

Consider a valve manifold as illustrated in Figure~\ref{fig:manifold}. 
The following model is derived by using the tank models from above for Tanks 1 and 2, the single pipe model for each pipe-valve-pipe connection with adjusted friction factor, and the orifice equation for the entrances of the tanks. We provide more details in the derivation subsequently.
\begin{tcolorbox}[title=Valve manifold model]
Consider Assumptions \ref{ass:pipeFlow}--\ref{ass:tank} (for isothermal pipes and tanks) and approximate the entering mass flow into Tanks~1 and 2 as an isentropic expansion described by orifice equation \eqref{eq:q-orifice}. Further, let each pipe-valve-pipe connection be approximated by single pipe models with adjusted friction factors. Let the state, output and input vectors be defined as 
\vspace{-5pt}
\begin{align*}
x &= \begin{bmatrix}
p_{T_1}&p_{1,r}& q_{1,\ell}&p_{2,r}&q_{2,\ell}&p_{T_2}
\end{bmatrix}^\top,\\
y &= \begin{bmatrix}
p_{T_2}&q_{T_1,\ell}
\end{bmatrix}^\top,\\
u&=\begin{bmatrix}
p_{T_1,\ell}&q_{T_2,r}
\end{bmatrix}^\top,
\end{align*}
where subscript $i=\{1,2\}$ refers to middle section $P_{ia}-V_i-P_{ib}$ and other terms as shown in Figure \ref{fig:manifold}.
 Then the overall dynamics are described by
{
\begin{subequations}
\begin{align*}
\dot x&=
\begin{multlined}[t]
\begin{bmatrix}
\alpha_{T_1}\xi_{T_1}&0&-\alpha_{T_1}&0&-\alpha_{T_1}&0\\
0 &\alpha_1\zeta_{T_2}&-\alpha_1&0&0&\alpha_1\xi_{T_2}\\
\kappa_1&\beta_1&\gamma_1&0&0&0\\
0&0&0&\alpha_2\zeta_{T_2}&-\alpha_2&\alpha_2\xi_{T_2}\\
\kappa_2&0&0&\beta_2&\gamma_2&0\\
0&\alpha_{T_2}\zeta_{T_2}&0&\alpha_{T_2}\zeta_{T_2}&0&2\alpha_{T_2}\xi_{T_2}
\end{bmatrix}x
+\begin{bmatrix}
\alpha_{T_1}\zeta_{T_1}&0&0&0&0&0\\
0&0&0&0&0&-\alpha_{T_2}
\end{bmatrix}^\top u,\end{multlined}\\
y& = \begin{bmatrix}
0 & 0 & 0 & 0 & 0 & 1\\
\xi_{T_1} & 0 & 0 & 0 & 0 & 0 
\end{bmatrix}x+\begin{bmatrix}
0&0\\ \zeta_{T_1}&0
\end{bmatrix}u,
\end{align*}
\end{subequations}
}
where $\xi_{T_1}=\xi_{T_1}(\bar p_{T_1,\ell},\bar p_{T_1})$ from \eqref{eq:orifice_lin_pr} and $\zeta_{T_1}=\zeta_{T_1}(\bar p_{T_1,\ell},\bar p_{T_1})$ from \eqref{eq:orifice_lin_pl} are related to the linearized orifice equation for the entrance into Tank 1, and accordingly for Tank~2. The parameters $\DpCoeff{i},\DqCoeffpr{i},\DqCoeffpl{i}$ and $\DqCoeffql{i}, i=\{1,2,3\}$ are as defined in \eqref{eq:pipe_coeff} for pipes, whereas $\alpha_{T_i}=\frac{R_sz_0T_0}{V_{T_i}},i=\{1,2\}$.
\end{tcolorbox}

\subsection{Derivation}
The model is based on the assumptions of each component, stated in the corresponding sections above. The primary motivation for including the orifice model for the entrance but not the exit is to obtain compatible boundary conditions. Our focus here is application and control orientation; an experimental validation is out of the scope of this work. {Without loss of generality}, we assume a positive flow, two middle sections, and that all outlets of Tank~1 and inlets of Tank~2 are of identical geometry, resp.
\paragraph*{Tanks}
For Tank~1, \eqref{eq:p_tank_iso} from the tank model above yields
\begin{align}
\dot p_{T_1}&=\alpha_{T_1}(q_{T_1,\ell}-q_{T_1,1,r}-q_{T_1,2,r}),\label{eq:T1-com}
\end{align}
where $q_{T_1,\ell}$ is the mass flow at the inlet of Tank~1 and $q_{T_1,i,r}$ is the mass flow at its $i^{\text{th}}$ outlet. We interpret the tank entrance as an orifice, and as in \eqref{eq:orifice_lin_pr} and \eqref{eq:orifice_lin_pl} define
{\small
\begin{align}
\frac{\partial g_{o}(p_{T_1},p_{T_1,\ell})}{\partial p_{T_1}}\doteq \xi_{T_1}, \quad \frac{\partial g_{o}(p_{T_1},p_{T_1,\ell})}{\partial p_{T_1,\ell}}\doteq \zeta_{T_1}.\label{eq:coeff_tank}
\end{align}}

Therefore, with \eqref{eq:T1-com} -- \eqref{eq:coeff_tank}, we obtain a model for Tank~1:
\begin{align*}
x&=\begin{bmatrix}
p_{T_1}
\end{bmatrix},\\
y&=\begin{bmatrix}
p_{T_1}&q_{T_1,\ell}
\end{bmatrix}^\top,\\
u&=\begin{bmatrix}
p_{T_1,\ell}&q_{T_1,1,r}&q_{T_1,2,r},
\end{bmatrix}^\top
\end{align*} with dynamics
{\small
\begin{subequations}\label{eq:tank1_model}
\begin{align}
\dot x&=\begin{bmatrix}
\alpha_{T_1}\xi_{T_1}
\end{bmatrix}x+\begin{bmatrix}
\alpha_{T_1}\zeta_{T_1} & -\alpha_{T_1} & -\alpha_{T_1}
\end{bmatrix}u,\\
 y&=\begin{bmatrix}
1 \\ \xi_{T_1}
\end{bmatrix}x+\begin{bmatrix}
0&0&0\\ \zeta_{T_1}&0 &0
\end{bmatrix}u.
\end{align}
\end{subequations}
}
Similarly, for Tank~2 we have:
\begin{align*}
x&=\begin{bmatrix}
p_{T_2}
\end{bmatrix},\\
y&=\begin{bmatrix}
p_{T_2}&q_{T_2,1,\ell}&q_{T_2,2,\ell}
\end{bmatrix}^\top\\
u&=\begin{bmatrix}
p_{T_2,1,\ell}&p_{T_2,2,\ell}&q_{T_2,r}
\end{bmatrix}^\top,
\end{align*}
with dynamics
{\small
\begin{subequations}\label{eq:tank2_model}
\begin{align}
\dot x&=\alpha_{T_2}(\xi_{T_2}+\xi_{T_2})
x+\begin{bmatrix}
\alpha_{T_2}\zeta_{T_2} & \alpha_{T_2}\zeta_{T_2} & -\alpha_{T_2}
\end{bmatrix}u,\\
 y&=\begin{bmatrix}
1 \\ \xi_{T_2}\\\xi_{T_2}
\end{bmatrix}x+\begin{bmatrix}
0&0&0\\ \zeta_{T_2}&0 &0\\ 0&\zeta_{T_2}&0
\end{bmatrix}u.
\end{align}
\end{subequations}
}
\paragraph*{Composite valve manifold model}
The state-space models for Tanks~1 and 2 in \eqref{eq:tank1_model} and \eqref{eq:tank2_model} as well as the single pipe model from \eqref{eq:linODEs} for each connection, $P_{1a}-V_1-P_{1b}$ and $P_{2a}-V_2-P_{2b}$, resp., share the following boundary conditions,
\vspace{-10pt}
\begin{align*}
p_{T_1,i,r}=p_{ia,\ell},\quad p_{T_2,i,\ell}=p_{ib,\ell}\\
q_{T_1,i,r}=q_{ia,\ell},\quad q_{T_2,i,\ell}=q_{ib,\ell},
\end{align*}
for $i=\{1,2\}$. Connecting tank and pipe models accordingly generates the  model stated above.
{
\section{Compressor}\label{sec:compressor}
Similar to the valve modeling, we present a static gain model as well as a more complex dynamic model for the compressor.
\subsection{Static model}
\begin{tcolorbox}[title=Static compressor model]
For the static gain model let the states and inputs be :
\begin{align}
x=\begin{bmatrix}
p_r&q_\ell
\end{bmatrix}^\top,\quad u=\begin{bmatrix}
p_\ell & q_r
\end{bmatrix}^\top,
\end{align}
where the suction side corresponds to subscript $\ell$ and discharge side to subscript $r$.
We define the static relationship
\begin{align*}
x=\begin{bmatrix}
k_c&0\\0 & 1
\end{bmatrix}u,
\end{align*}
where $k_c>1$. 
\end{tcolorbox}
Conservation of mass is immediate.
\subsection{Dynamic model}
\begin{figure}[ht]
    \centering
   {\includegraphics{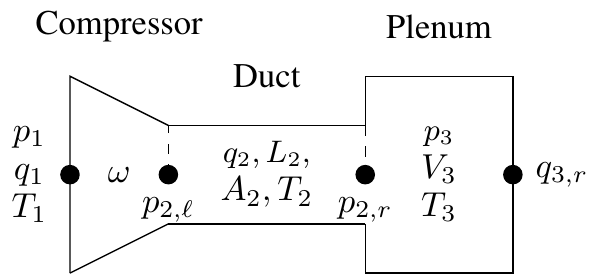}}
 \caption{Compressor system consisting of compressor, duct and plenum. The compression is isentropic, whereas duct and plenum are isothermal. We consider boundary conditions $p_1,q_{3,r}$ and $T_1$ as known. Variables at internal boundaries are continuous.}
    \label{fig:compressor}
\end{figure}
Consider a compressor system as shown in Figure \ref{fig:compressor}, which consists of a compressor, a duct and a plenum. Similar to \cite{greitzer1976} and \cite{egeland2002}, we model each of these elements separately and then connect them via boundary conditions, akin to the valve manifold model above.
\begin{tcolorbox}[title=Dynamic compressor model]
Consider Assumptions~\ref{ass:pipeFlow} (for duct), \ref{ass:tank} (for isothermal plenum) and \ref{ass:compr} (for compression) below.
Define state, input and output vectors as
\begin{align}
x&= \begin{bmatrix}
 p_3&
 q_2
\end{bmatrix}^\top,\\
u &= \begin{bmatrix}
p_1&q_{3,r}&T_1& \omega
\end{bmatrix} ^\top,\\
y &= \begin{bmatrix}
p_3&q_{2} & T_3
\end{bmatrix} ^\top,
\end{align}
with terms as shown in Figure \ref{fig:compressor}, and $\omega$ the compressor speed. The the nonlinear dynamics are
\begin{align}
\dot x& = \begin{bmatrix}
\frac{R_sT_{2}z_0}{V_p}(q_{2}-q_{3,r})\\
\frac{A_2}{L_2}(\Phi(q_2,\omega)-p_3)
\end{bmatrix}\\
y&=\begin{bmatrix}
x\\
T_1\left(\frac{\Phi(q_2,\omega)}{p_1}\right)^{(\eta-1)/\eta}
\end{bmatrix},
\end{align}
where $\Phi(q_2,\omega)=p_{2,\ell}$ is the suction pressure after the isentropic compression modeled as a static map. A linear state space model can be derived by linearization.
\end{tcolorbox}
\subsubsection{Assumptions}
\begin{ass}\label{ass:compr} Suppose:
\begin{enumerate}[label=\ref{ass:compr}(\roman*)]
\item The compression is isentropic.
\item The discharge pressure of the compressor, $p_{2,\ell}$, is defined by a static relation between speed, $\omega$, suction pressure, $p_1$, and mass flow, $q_2$: $\Phi(q_2,\omega): \;\R_{>0}\times \R_{>0}\to\R_{>0}$.
\item The friction in the duct is negligible.
\item The temperature change in the duct is negligible.
\item The gas in the duct and plenum is perfectly mixed.
\end{enumerate}

\end{ass}

\subsubsection{Derivation}
For the plenum we obtain from the isothermal tank model in \eqref{eq:p_tank_iso},
\begin{align}
\dot p_3 = \frac{R_sT_{2}z_0}{V_p}(q_{2}-q_{3,r}).\label{eq:p3dot}
\end{align}
For the duct, by \eqref{eq:dqldt_iso} and boundary condition $p_{2,r}=p_3$,
\begin{align}
\dot q_2 = \frac{A_2}{L_2}(p_{2,\ell}-p_3).
\end{align}
The suction pressure after the isentropic compression is modeled as a static map, $p_{2.\ell}=\Phi(q_2,\omega)$, with compressor speed $\omega$. The mapping is usually based on empirical formulas \cite{rennels2012pipe}, or compressor geometry \cite[Ch. 13]{egeland2002}. It follows that
\begin{align}
\dot q_2 = \frac{A_2}{L_2}(\Phi(q_2,\omega)-p_3).\label{eq:q2dot}
\end{align}
By the hypothesis of isentropic compression and isothermal duct and plenum,
\begin{align}
T_3=T_1\left(\frac{p_{2,\ell}}{p_1}\right)^{(\eta-1)/\eta}=T_1\left(\frac{\Phi(q_2,\omega)}{p_1}\right)^{(\eta-1)/\eta},\label{eq:T3}
\end{align}
with isentropic coefficient $\eta$. The model above immediately follows. Note that conservation of mass is immediate: by hypothesis and \eqref{eq:p3dot}, $\bar q_{1}=\bar q_{2}=\bar q_{3,r}$. 
}
{
\section{Heat exchanger}
We provide a heat exhanger model based on the non-isothermal one-dimensional pipe model presented in \cite{sven_bob_rob_gas}. The assumptions are therefore similar to those for the pipe in Section \ref{subsec:pipe_ass}. Given the fact that under usual operating conditions no phase {change} occurs {within the gas} our model is similar to that in \cite{Rasmussen2004} in the context of air conditioning systems. While we provide a nonlinear formulation, linearization around the nominal operating point directly leads to a linear state-space realization.

{Our rudimentary model replaces the heat exchanger with a single pipe, with the geometric and conductivity properties} accommodated by adjusting the heat transfer coefficient, $k_{\text{rad}}$, and friction factor, $\lambda${, to parametrize the exchanger design and nominal operating point}. The temperature output of the model can be used to set the temperature parameter for {subsequent downstream} isothermal {(or other)} pipe models.
\begin{tcolorbox}[title=Heat exchanger]
Considering Assumption \ref{ass:pipeFlow}, if the heat flux {is} dominated by radial conduction we define the state and input vectors {to be}
\begin{align*}
x&=\begin{bmatrix}
p_r & q_\ell & T_r
\end{bmatrix},\\
u &= \begin{bmatrix}
p_\ell & q_r & T_\ell
\end{bmatrix}.
\end{align*}
Then, the nonlinear dynamics of the heat exchanger are governed by
\begin{subequations}\label{eq:3d_model_eq}
\begin{align}
\dot{p}_r&=\begin{multlined}[t] \frac{R_sz_0}{Ac_v}\left[k_\text{rad}  \pi D_o (T_\text{amb}-T_r)
-\frac{q_r-q_\ell}{L}T_r\left(c_v+R_sz_0\right)+\frac{p_r-p_\ell}{L} \frac{R_sz_0T_rq_r}{p_r}\right.\\
%
\left.-\frac{T_r-T_\ell}{L}q_r\left(c_v+R_sz_0\right)+\frac{\lambda R_s^2z_0^2T_r^2q_r^2|q_r|}{2DA^2p_r^2}\right]
\end{multlined}\label{eq:ode_p_noniso}
\\
\dot{q}_\ell&=-A\frac{p_r-p_\ell}{L}-\frac{\lambda R_sT_\ell z_0}{2DA}\frac{q_\ell\vert q_\ell\vert}{p_\ell}-\frac{Ag}{R_sT_\ell z_0} \frac{d h}{dx} p_\ell\label{eq:ode_q_noniso}\\
\dot{T}_r&=\begin{multlined}[t] \frac{R_sz_0T_r}{Ac_vp_r}\left[k_\text{rad}  \pi D_o (T_\text{amb}-T_r)-\frac{q_r-q_\ell}{L}T_rR_sz_0
+\frac{p_r-p_\ell}{L} \frac{R_sz_0T_rq_r}{p_r}-\frac{T_r-T_\ell}{L}q_r\left(c_v+R_sz_0\right)\right.\\
%
\left.+\frac{\lambda R_s^2z_0^2 T_r^2q_r^2|q_r|}{2DA^2p_r^2}\right]\end{multlined}\label{eq:ode_T_noniso}
\end{align}
\end{subequations}
\end{tcolorbox}

\subsection{Derivation}
{This is the spatially discretized model of non-isothermal pipe flow developed} in \cite{sven_bob_rob_gas}. {As remarked above, the heat transfer coefficient, $k_\text{rad}$, is deliberately non-negligible and the friction factor, $\lambda$, can be significant depending on exchanger geometry.}
}
\section{Interconnections}
Suppose each component is described by a state space model with signals describing pressures, $p$, and mass flows, $q$. In this section we show how these component models can be interconnected to build an entire network model. Towards this goal, we recall interconnection rules from \cite{sven_bob_control} which translate allowable boundary conditions to signal flows.

\subsection{Directed pipe connections and `ports'}
The signal flow graph models derived later have directions associated with each signal. Thus, $p_\ell$ and $q_r$ are input signals, indicating that they are specified from outside the component, and $p_r$ and $q_\ell$ are output signals, meaning that they are determined by the component itself and the input signals. The spatially localized connection sites, however, possess one input signal and one output signal, namely $(p_\ell,q_\ell)$ at the left end and $(p_r,q_r)$ at the right. We further appropriate circuit terminology and identify two distinct location types, which we term \textit{ports}. As in \cite{benner2019}, every element in our interconnected system presents signal interfaces to other elements and to the outside world.

\begin{defin}[Ports]
\begin{description}
\item
\item[$\boldsymbol p$-port]  of a component possesses two signals: an input pressure signal $p_\ell$ and an output flow signal $q_\ell$.
\item[$\boldsymbol q$-port] of a component possesses two signals: an input flow signal $q_r$ and an output pressure signal $p_r$.
\end{description}
\end{defin}

Internal series connection of two components, 1 and 2, will involve the cascading of signals $p_{2,\ell}=p_{1,r}$ and $q_{1,r}=q_{2,\ell}$ at the junction point. This describes a $p$-port to $q$-port connection. Likewise, connection to the outside of the network must respect the type and causality of the signals. These rules are specified below.

\subsection{Interconnection rules}
\begin{tcolorbox}[title=Interconnection rules]
\begin{enumerate}[label=\Roman*.]
\item \vskip -2mm 
Connections are permitted only between:
\begin{enumerate}[label=\roman*.]
\item a $p$-port and a $q$-port, or
\item a $p$-port and an external pressure source/input signal plus an external flow sink/output signal, or
\item a $q$-port and an external flow source/input signal plus an external pressure sink/output signal.
\end{enumerate}
\item Pressure input signals must connect to pressure output signals, and flow input signals must connect to flow output signals. 
\item Connection of one variable of a port requires connection of the other.
\item All ports must be connected and algebraic loops avoided.
\end{enumerate}
\end{tcolorbox}

These rules conform to the connections examined in \cite{sven_bob_control} to formulate the systematic interconnection of state-space models. Each component model possesses input and output signals and, in Proposition~3 \cite{sven_bob_rob_gas}, it is shown how a (possibly non-minimal) state-space realization of the interconnection of gas system elements can be directly constructed with the above rules. This construction replaces and extends the graph-theoretic DAE methods of \cite{benner2019} and yields a new input-output transfer function satisfying Mason's Gain Formula, \cite[Proposition~4]{sven_bob_rob_gas}. 

\subsection{Matrix methodology}
Per \cite{sven_bob_rob_gas}, we begin by stacking the state-space models of the individual network components:
\begin{align}\label{eq:totalSystem}
\dot{ x}=A x+B w,\quad  y=C x+D w
\end{align}
where
\begin{subequations}\label{eq:matrices_stacked_up}
\begin{align}
A&=\blkdiag(A^{(1)}, A^{(2)}, \dots, A^{(N)}),\\
B&=\blkdiag(B^{(1)}, B^{(2)}, \dots, B^{(N)}),\\
C&=\blkdiag(C^{(1)}, C^{(2)}, \dots, C^{(N)}),\\
D&=\blkdiag(D^{(1)}, D^{(2)}, \dots, D^{(N)}),
\end{align}
\end{subequations}
with superscript $(i),i=\{1,\dots,N\},$ denoting the respective matrix of component $i$, and $x\in\R^{n_x},w\in\R^{n_w}$ and $y\in\R^{n_y}$.
Interconnections and external sources $u\in\R^{n_u}$ {and sinks $z\in\R^{n_z}$} are described by 
\begin{align}\label{eq:FG_network}
 w = F y+G u,\;\;\; z = Hx+J u,
\end{align}
with {structured matrices} $[F,G,H,J]$ {with 0-1 elements}:
\begin{align}\label{eq:connection_matrices}
[F]_{i,j}&=\begin{cases}
1, & \text{if } [ y]_j=[ w]_i,\\
0, & \text{otherwise}.
\end{cases}
\end{align}
\begin{tcolorbox}[title=Matrix interconnection methodology]
A (perhaps non-minimal) state-space realization of \eqref{eq:totalSystem} and \eqref{eq:FG_network} is given by
\begin{align}\label{eq:matrix_framework}
x=\bar A  x+\bar B  u, \quad  y=\bar C  x+\bar D  u,
\end{align}
where
\begin{align*}
\bar A&=A+BF(I-DF)^{-1}C, \quad \bar C=(I-DF)^{-1}C,\\
\bar B&=B\left[I+F(I-DF)^{-1}D\right]G,\quad \bar D=(I-DF)^{-1}DG.
\end{align*}
\end{tcolorbox}
We next show an example applying the methodology.
\subsubsection{Example: two pipes in series}
Consider two pipes in series, 1 and 2, with aggregated input and state/output vectors,
\begin{align}
x =y=\begin{bmatrix}
p_{1,r}\\q_{1,\ell}\\p_{2,r}\\q_{2,\ell}
\end{bmatrix}, \quad w=\begin{bmatrix}
p_{1,\ell}\\q_{1,r}\\p_{2,\ell}\\q_{2,r}
\end{bmatrix},
\end{align}
We further define the sources and sinks:
\begin{align}
u = \begin{bmatrix}
p_{1,\ell}\\q_{2,r}
\end{bmatrix}, \quad z=\begin{bmatrix}
p_{2,r}\\
q_{1,\ell}
\end{bmatrix}.
\end{align}
As $p_{1,r}=p_{2,\ell}$ and $q_{1,r}=q_{2,\ell}$, this leads to the following matrices:
\begin{align}
F&=\begin{bmatrix}
0&0&0&0\\
0 &0& 0& 1 \\
1 & 0 & 0 & 0\\
0&0&0&0
\end{bmatrix},\quad 
G=\begin{bmatrix}
1&0\\
0 &0 \\
0 & 0 \\
0&1
\end{bmatrix}, \\
H&=\begin{bmatrix}
0&0&1&0\\
0 &1&0&0 \\
\end{bmatrix},\quad  J=\zero.
\end{align}
The aggregate system can directly be implemented in Matlab using \eqref{eq:matrix_framework}. 
\subsubsection{Matlab example: {vented} gas loop}
\begin{figure}[ht!]
    \centering
   \includegraphics{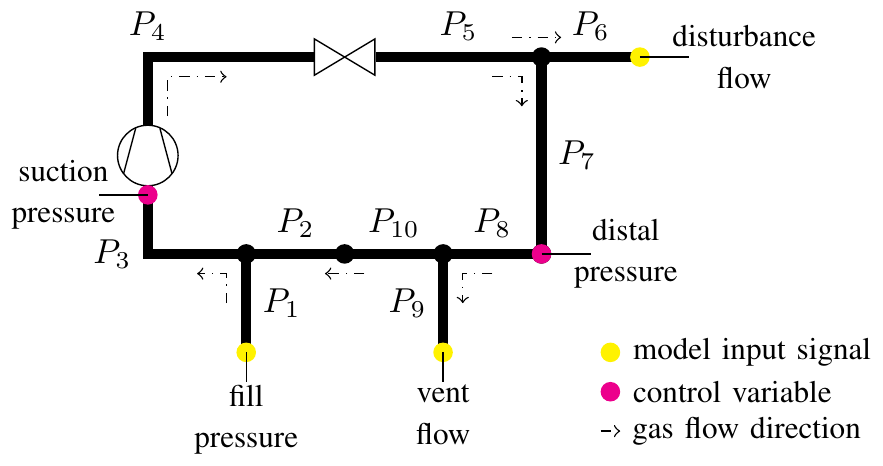}
  \caption{Pipe network with compressor and valve $\bowtie$. In process control parlance, the fill pressure and vent flow are manipulated variables, the suction and distal pressures are controlled variables, and the flow from $P_6$ is a disturbance signal.}
    \label{fig:control_loop}
\end{figure}

We recall the example from \cite{sven_bob_rob_gas} in Figure \ref{fig:control_loop}.

The gas flows clockwise, entering through pipe $P_1$ and exiting through pipes $P_6$  and $P_9$. The compressor and valve, whose corresponding variables are respectively labeled by subscripts $c$ and $v$, are modeled as static gains as in Sections~\ref{sec:valve} and \ref{sec:compressor},
\begin{align*}
D_c=\begin{bmatrix}
k_c&0\\0 & 1
\end{bmatrix},\quad D_v=\begin{bmatrix}
k_v&0\\0 & 1
\end{bmatrix},
\end{align*}
where $k_c=4$ and $k_v=0.8$. Further, pipes $(P_1,P_2,P_3)$ are modeled as a joint, as in Subsection~\ref{subsec:joint}, and $(P_5,P_6,P_7)$ and $(P_8,P_9,P_{10})$ as branches, as in Section~\ref{sec:branch}.
Composing the system according to \eqref{eq:matrices_stacked_up}, results in the component input vector,
\begin{align*}
 w&=\left[\begin{matrix}
p_{1,\ell}& p_{2,\ell} & q_{3,r} & p_{c,\ell} & q_{c,r} & p_{4,\ell} & q_{4,r} \end{matrix}\right.
\left.\begin{matrix} p_{v,\ell} & q_{v,r} & p_{5,\ell} & q_{6,r} & q_{7,r} & p_{8,\ell}q_{9,r} & q_{10,r}
\end{matrix}\right]^\top,
\end{align*}
and the total output vector,
\begin{align*}
 y&=\left[\begin{matrix}
p_{3,r}& q_{1,\ell} & q_{2,\ell} & p_{c,r} & q_{c,\ell} & p_{4,r} & q_{4,\ell} \end{matrix}\right.
 \left.\begin{matrix} p_{v,r} & q_{v,\ell} & p_{6,r} & p_{7,r} & q_{5,\ell} & p_{9,r}&p_{10,r} & q_{8,\ell}
\end{matrix}\right]^\top.
\end{align*}
The external sources are
\begin{align*}
\bar u_t&=\begin{bmatrix}
p_{1,\ell} & q_{6,r} & q_{9,r}
\end{bmatrix}^\top.
\end{align*}
\paragraph*{Matlab code}\mbox{}\\
\begin{lstlisting}[style=Matlab-editor]
%% Parameters
clear all
close all

pbar = 2.54e6; % nominal pressure Pa
qbar = 15.44; % nominal flow Kg/s
Re = 1.0901e7; % Reynolds number
T = 300; % nominal temperature K
z = 1; % gas compression factor
Rs = 518.28; % specific gas constant JK^{-1}mol^{-1} methane
lambda = 0.0113; % friction factor
gamma = Rs*T; % constant gas state
mu = 1.31; % from https://www.engineeringtoolbox.com/methane-d_1420.html

%% System J123 join of three 10m pipes
La = 10; % m
Da = 27.25*2.54/100; % m
Aa = pi*Da*Da/4;

alfa = gamma*z/Aa/La;
beta = Aa/La;
kappa = lambda*gamma*z*qbar/Da/Aa/pbar;

Lb = 10; % m
Db = 17.25*2.54/100; % m
Ab = pi*Db*Db/4;

alfb = gamma*z/Ab/Lb;
betb = Ab/Lb;
kappb = lambda*gamma*z*qbar/Db/Ab/pbar;

Lc = 10; % m
Dc = 27.25*2.54/100; % m
Ac = pi*Dc*Dc/4;

alfc = gamma*z/Ac/Lc;
betc = Ac/Lc;
kappc = lambda*gamma*z*qbar/Dc/Ac/pbar;

gammab = alfa/(alfa+alfb); % splitting ratio for joining problem

F123 = [0 alfa*(1-gammab) alfa*(1-gammab) 0 -alfa*(1-gammab);
    -beta -kappa 0 0 0;
    -betb 0 -kappb 0 0;
    0 0 0 0 alfc;
    betc 0 0 -betc -kappc];
G123 = [0 0 0;beta 0 0;0 betb 0;0 0 -alfc;0 0 0];
H123 = [0 0 0 1 0;0 1 0 0 0;0 0 1 0 0];
J123 = zeros(3,3);

%% Systems Br567 and Br8910 identical branching systems
F567 = [0 alfa 0 -alfa 0 -alfa;
    -beta -kappa 0 0 0 0;
    0 0 0 alfb 0 0;
    betb 0 -betb -kappb 0 0;
    0 0 0 0 0 alfc;
    betc 0 0 0 -betc -kappc];
G567 = [0 0 0; beta 0 0; 0 -alfb 0;0 0 0;0 0 -alfc;0 0 0];
H567 = [0 0 1 0 0 0;0 0 0 0 1 0;0 1 0 0 0 0];
J567 = zeros(3,3);
F8910=F567; G8910=G567; H8910=H567; J8910=J567;

%% Single pipe 4
F4 = [0 alfa;-beta -kappa];
G4 = [0 -alfa;beta 0];
H4 = eye(2);
J4 = zeros(2,2);

%% System C compressor
kC = 4;
FC = zeros(0);
GC = zeros(0,2);
HC = zeros(2,0);
JC = diag([kC,1]);

%% System resistor
kR = 0.8;
FR = zeros(0);
GR = zeros(0,2);
HR = zeros(2,0);
JR = diag([kR,1]);

%% Compose interconnected system

% stack up components
A = blkdiag(F123,FC,F4,FR,F567,F8910);
B = blkdiag(G123,GC,G4,GR,G567,G8910);
C = blkdiag(H123,HC,H4,HR,H567,H8910);
D = blkdiag(J123,JC,J4,JR,J567,J8910);

% Internal connections
F = zeros(15,15);
F(2,14)=1; F(3,5)=1; F(4,1)=1; F(5,7)=1; F(6,4)=1; F(7,9)=1;
F(8,6)=1; F(9,12)=1; F(10,8)=1; F(12,15)=1; F(13,11)=1; F(15,3)=1;

% External sources
G = zeros(15,3);
G(1,1)=1; G(11,3)=1; G(14,2)=1; 

% Interconnected systems
Abar = A+B*F/(eye(15)-D*F)*C;
Bbar = B*(eye(15)+F/(eye(15)-D*F)*D)*G;
Cbar = (eye(15)-D*F)\C;
Dbar = (eye(15)-D*F)\D*G;

%% State space model
complete_sys = ss(Abar, Bbar, Cbar, Dbar);
\end{lstlisting}

\subsection{Matlab's connect function}
As an alternative to the matrix methodology we can use the \verb|connect| function in Matlab. This is best explained by the example below. For each component we name the input and output signals, respecting the boundary conditions. We then define the input and output signal names of the aggregate system and call the \verb|connect| function.
\subsubsection{Matlab example: two pipes in series}

\begin{lstlisting}[style=Matlab-editor]
% sys1 - Pipe 1
% sys2 - Pipe 2

% Pipe 1
sys1.InputName = {'p1l', 'q2l'};  % q1r = q2l
sys1.OutputName = {'p1r', 'q1l'};

% Pipe 2
sys2.InputName = {'p1r', 'q2r'};  % p1r = p2l
sys2.OutputName = {'p2r', 'q2l'};

% Connect pipes
sys_inputs = {'p1l', 'q2r'};
sys_outputs = {'p2r', 'q1l'};
sys = connect(sys1,sys2, sys_inputs, sys_outputs);
\end{lstlisting}
\subsubsection{Matlab example: {vented} gas loop}
\begin{lstlisting}[style=Matlab-editor]
%% Parameters
clear all
close all

pbar = 2.54e6; % nominal pressure Pa
qbar = 15.44; % nominal flow Kg/s
Re = 1.0901e7; % Reynolds number
T = 300; % nominal temperature K
z = 1; % gas compression factor
Rs = 518.28; % specific gas constant JK^{-1}mol^{-1} methane
lambda = 0.0113; % friction factor
gamma = Rs*T; % constant gas state
mu = 1.31; % from https://www.engineeringtoolbox.com/methane-d_1420.html

%% System J123 join of three 10m pipes
La = 10; % m
Da = 27.25*2.54/100; % m
Aa = pi*Da*Da/4;

alfa = gamma*z/Aa/La;
beta = Aa/La;
kappa = lambda*gamma*z*qbar/Da/Aa/pbar;

Lb = 10; % m
Db = 17.25*2.54/100; % m
Ab = pi*Db*Db/4;

alfb = gamma*z/Ab/Lb;
betb = Ab/Lb;
kappb = lambda*gamma*z*qbar/Db/Ab/pbar;

Lc = 10; % m
Dc = 27.25*2.54/100; % m
Ac = pi*Dc*Dc/4;

alfc = gamma*z/Ac/Lc;
betc = Ac/Lc;
kappc = lambda*gamma*z*qbar/Dc/Ac/pbar;

gammab = alfa/(alfa+alfb); % splitting ratio for joining problem

F123 = [0 alfa*(1-gammab) alfa*(1-gammab) 0 -alfa*(1-gammab);
    -beta -kappa 0 0 0;
    -betb 0 -kappb 0 0;
    0 0 0 0 alfc;
    betc 0 0 -betc -kappc];
G123 = [0 0 0;beta 0 0;0 betb 0;0 0 -alfc;0 0 0];
H123 = [0 0 0 1 0;0 1 0 0 0;0 0 1 0 0];
J123 = zeros(3,3);
sys123 = ss(F123,G123,H123,J123);
sys123.InputName = {'p1l','p10r','qCl'};
sys123.OutputName = {'p3r','q1l','q2l'};


%% Systems Br567 and Br8910 identical branching systems
F567 = [0 alfa 0 -alfa 0 -alfa;
    -beta -kappa 0 0 0 0;
    0 0 0 alfb 0 0;
    betb 0 -betb -kappb 0 0;
    0 0 0 0 0 alfc;
    betc 0 0 0 -betc -kappc];
G567 = [0 0 0; beta 0 0; 0 -alfb 0;0 0 0;0 0 -alfc;0 0 0];
H567 = [0 0 1 0 0 0;0 0 0 0 1 0;0 1 0 0 0 0];
J567 = zeros(3,3);
sys567 = ss(F567,G567,H567,J567);
sys567.InputName = {'pVr','q6r','q8l'};
sys567.OutputName = {'p6r','p7r','q5l'};
sys8910 = sys567;
sys8910.InputName = {'p7r', 'q9r', 'q2l'};
sys8910.OutputName = {'p9r', 'p10r', 'q8l'};

%% Single pipe 4
F4 = [0 alfa;-beta -kappa];
G4 = [0 -alfa;beta 0];
H4 = eye(2);
J4 = zeros(2,2);
sys4 = ss(F4,G4,H4,J4);
sys4.InputName = {'pCr', 'qVl'};
sys4.OutputName = {'p4r', 'q4l'};

%% System C compressor
kC = 4;
FC = zeros(0);
GC = zeros(0,2);
HC = zeros(2,0);
JC = diag([kC,1]);
sysC = ss(FC,GC,HC,JC);
sysC.InputName = {'p3r', 'q4l'};
sysC.OutputName = {'pCr', 'qCl'};

%% System resistor
kR = 0.8;
FR = zeros(0);
GR = zeros(0,2);
HR = zeros(2,0);
JR = diag([kR,1]);
sysR = ss(FR,GR,HR,JR);
sysR.InputName = {'p4r', 'q5l'};
sysR.OutputName = {'pVr', 'qVl'};

%% Compose interconnected system
inputs = {'p1l', 'q6r', 'q9r'};
outputs = {'p3r','q1l','q2l','p6r','p7r','q5l','p9r','p10r','q8l','p4r','q4l','pVr', 'qVl'};
complete_sys = connect(sysR,sysC,sys4,sys123,sys567,sys8910, inputs, outputs);
\end{lstlisting}

\section{Conclusion}
We derived composite state-space models for the transient dynamics of gas flow through intersecting pipe geometries, valves, compressors, and valve manifolds that are well-suited candidates for model-based control design. They also capture conservation of mass at steady state by subsuming algebraic constraints that would otherwise appear as part of a system of DAEs. Additionally, we provide examples of how to use the matrix methodology introduced in \cite{sven_bob_rob_gas} in Matlab and alternatively show how to use Matlab's \verb|connect| function.

\section*{Acknowledgement}
This research was supported by funding from Solar Turbines {Incorporated}.

\bibliographystyle{plain}
\bibliography{bib_all.bib,ref2.bib,/Users/sven/Documents/MEGA/Uni/Latex_ressources/bib_all.bib}

\end{document}